\documentclass[12pt]{article}

\usepackage[dvips]{graphicx}
\usepackage{wrapfig}
\usepackage{amssymb}
\usepackage{amsmath}
\usepackage{cases}

\def\IN{\mathbb {N}}
\def\IZ{\mathbb {Z}}
\def\IQ{\mathbb {Q}}
\def\IR{\mathbb {R}}
\def\IC{\mathbb {C}}
\def\IP{\mathbb {P}}
\def\IF{\mathbb {F}}

\renewcommand{\thefootnote}{\fnsymbol{footnote}}

\newdimen\tableauside\tableauside=1.0ex
\newdimen\tableaurule\tableaurule=0.4pt
\newdimen\tableaustep
\def\phantomhrule#1{\hbox{\vbox to0pt{\hrule height\tableaurule width#1\vss}}}
\def\phantomvrule#1{\vbox{\hbox to0pt{\vrule width\tableaurule height#1\hss}}}
\def\sqr{\vbox{%
  \phantomhrule\tableaustep
  \hbox{\phantomvrule\tableaustep\kern\tableaustep\phantomvrule\tableaustep}%
  \hbox{\vbox{\phantomhrule\tableauside}\kern-\tableaurule}}}
\def\squares#1{\hbox{\count0=#1\noindent\loop\sqr
  \advance\count0 by-1 \ifnum\count0>0\repeat}}
\def\tableau#1{\vcenter{\offinterlineskip
  \tableaustep=\tableauside\advance\tableaustep by-\tableaurule
  \kern\normallineskip\hbox
    {\kern\normallineskip\vbox
      {\gettableau#1 0 }%
     \kern\normallineskip\kern\tableaurule}%
  \kern\normallineskip\kern\tableaurule}}
\def\gettableau#1 {\ifnum#1=0\let\next=\null\else
  \squares{#1}\let\next=\gettableau\fi\next}

\makeatletter
 \renewcommand{\theequation}{%
       \thesection.\arabic{equation}}
 \@addtoreset{equation}{section}
\makeatother

\makeatletter
\def\eqnarray{%
 \stepcounter{equation}%
 \let\@currentlabel=\theequation
 \global\@eqnswtrue
 \global\@eqcnt\z@
 \tabskip\@centering
 \let\\=\@eqncr
 $$\halign to \displaywidth\bgroup\@eqnsel\hskip\@centering
 $\displaystyle\tabskip\z@{##}$&\global\@eqcnt\@ne
 \hfil$\displaystyle{{}##{}}$\hfil
 &\global\@eqcnt\tw@$\displaystyle\tabskip\z@{##}$\hfil
 \tabskip\@centering&\llap{##}\tabskip\z@\cr}
\makeatother

\begin{document}

\begin{titlepage}

\begin{center}
\vspace*{1cm}
{\Large \bf
Stringy Instanton Counting and Topological Strings}
\lineskip .75em
\vskip 1.5cm
{\large Masahide Manabe\footnote[1]{masahidemanabe@gmail.com}}
\vskip 1.0em
{\it 
Faculty of Physics, University of Warsaw\\
ul. Pasteura 5, 02-093 Warsaw, Poland\\}
\end{center}
\vskip3cm

\begin{abstract}
We study the stringy instanton partition function of four dimensional ${\cal N}=2$ $U(N)$ supersymmetric gauge theory which was obtained by Bonelli et al in 2013. In type IIB string theory on ${\IC}^2\times T^*{\IP}^1\times {\IC}$, the stringy $U(N)$ instantons of charge $k$ are described by $k$ D1-branes wrapping around the ${\IP}^1$ bound to $N$ D5-branes on ${\IC}^2\times {\IP}^1$. The KK corrections induced by compactification of the ${\IP}^1$ give the stringy corrections. We find a relation between the stringy instanton partition function whose quantum stringy corrections have been removed and the K-theoretic instanton partition function, or by geometric engineering, the refined topological A-model partition function on a local toric Calabi-Yau threefold. We also study the quantum stringy corrections in the stringy instanton partition function which is not captured by the refined topological strings.
\end{abstract}
\end{titlepage}

\renewcommand{\thefootnote}{\arabic{footnote}} \setcounter{footnote}{0}

\section{Introduction}\label{sec:introduction}

In 1994, by Seiberg and Witten the exact prepotential ${\cal F}_0={\cal F}_0^{\mbox{\scriptsize pert}}+{\cal F}_0^{\mbox{\scriptsize inst}}$ in four dimensional ${\cal N}=2$ $SU(2)$ supersymmetric gauge theory was obtained \cite{Seiberg:1994rs,Seiberg:1994aj}, where ${\cal F}_0^{\mbox{\scriptsize pert}}$ and ${\cal F}_0^{\mbox{\scriptsize inst}}$ are the perturbative and instanton part of the prepotential, respectively. The prepotential is computed from a period of a two dimensional algebraic curve on the Coulomb branch which is called Seiberg-Witten curve. This exact result can be generalized to other gauge theories with $ADE$ gauge symmetries. By the compactification of type IIA strings on a local Calabi-Yau threefold given by ALE space fibration of $ADE$ type over ${\IP}^1$, one obtains ${\cal N}=2$ supersymmetry with  $ADE$ gauge symmetry in four dimensions. This realization is known as geometric engineering of gauge theory \cite{Klemm:1996bj, Katz:1996fh, Katz:1997eq}. Then the Seiberg-Witten curve is embedded into the mirror dual of the local Calabi-Yau threefold. In 2002, by Nekrasov the prepotential ${\cal F}_0$ was directly derived from path integral formulation using the localization technique \cite{Nekrasov:2002qd}. A necessary ingredient of this computation is to introduce the Omega background described by ${\IR}^4\cong {\IC}^2$ fibration over two dimensional torus $T^2$. The Omega background has two generators $\epsilon_{1}$ and $\epsilon_{2}$ of $T^2$. The instanton moduli space of the gauge theory can be described by the ADHM moduli space whose dynamical variables are given by matrices \cite{Atiyah:1978ri} (see (\ref{ADHM_moduli})). The computation of the path integral which gives the instanton partition function $Z^{\mbox{\scriptsize Nek}}(\epsilon_1,\epsilon_2)$ on the Omega background reduces to the computation of the equivariant volume of the ADHM moduli space whose IR behavior was regularized by $\epsilon_{1,2}$ \cite{Moore:1997dj, Moore:1998et, Nekrasov:2002qd}.\footnote{The instanton partition function $Z^{\mbox{\scriptsize Nek}}(\epsilon_1,\epsilon_2)$ also depends on the Coulomb moduli described by the Cartan subalgebra for gauge group, the dynamical scales in four dimensional gauge theory, and the masses of matters if the gauge theory is coupled to them. In this introduction we abbreviate these arguments for simplicity.} By the localization, the instanton partition function $Z^{\mbox{\scriptsize Nek}}(\epsilon_1,\epsilon_2)$ can be exactly computed, and one obtains the asymptotic expansion of the form
$$
\log Z^{\mbox{\scriptsize Nek}}(\epsilon_1,\epsilon_2)=\sum_{g,\ell=0}^{\infty}(\epsilon_1\epsilon_2)^{g-1}(\epsilon_1+\epsilon_2)^{\ell}{\cal F}_{g,\ell}^{\mbox{\scriptsize Nek}}.
$$
Here the leading term coincides with the instanton part of the prepotential ${\cal F}_{0,0}^{\mbox{\scriptsize Nek}}={\cal F}_0^{\mbox{\scriptsize inst}}$ \cite{Nekrasov:2002qd, Nakajima:2003pg, Nekrasov:2003rj, Braverman:2004cr}. The $SU(N)$ instanton partition function on the anti-self-dual Omega background $\epsilon_1=-\epsilon_2=\hbar$ can be also obtained from a geometric engineering limit of the topological A-model partition function on a local toric Calabi-Yau threefold given by ALE space fibration of $A_{N-1}$ type over ${\IP}^1$ \cite{Iqbal:2003ix, Iqbal:2003zz, Eguchi:2003sj, Hollowood:2003cv, Eguchi:2003it, Zhou:2003zp}. Here the topological A-model partition function is computed by the topological vertex formalism \cite{Iqbal:2002we, Aganagic:2003db}, and the parameter $\hbar$ is identified with the topological string coupling constant $g_s$. In \cite{Awata:2005fa, Iqbal:2007ii}, refinements of the topological vertex formalism were proposed so that the $SU(N)$ instanton partition function on the general Omega background is obtained \cite{Iqbal:2007ii, Taki:2007dh, Awata:2008ed}.

In 2012, the partition function of ${\cal N}=(2,2)$ gauged linear sigma model (GLSM) on $S^2$ was exactly computed \cite{Benini:2012ui,Doroud:2012xw}. By using this result, the stringy instanton partition function of four dimensional ${\cal N}=2$ $U(N)$ supersymmetric gauge theory was given in \cite{Bonelli:2013rja}. The GLSM, which gives the stringy instanton partition function, flows in the IR fixed point to an ${\cal N}=(2,2)$ non-linear sigma model (NLSM) whose target space is given by the ADHM moduli space. In \cite{Bonelli:2013rja}, it was shown that in the degenerate limit of the worldsheet $S^2$ the stringy instanton partition function yields the $U(N)$ instanton partition function $Z_N^{\mbox{\scriptsize Nek}}(\epsilon_1,\epsilon_2)$ (see (\ref{Nekrasov_in}) for the $k$-instanton sector).\footnote{The stringy instanton counting in this paper means the ``gauge theoretic'' instanton counting with the stringy corrections. This is different from ``stringy'' (or ``exotic'') instantons discussed in e.g. \cite{Blumenhagen:2006xt, Ibanez:2006da, Florea:2006si, Argurio:2007vqa, Bianchi:2007wy, Billo:2009di, Fucito:2009rs, Billo':2010bd, Ghorbani:2010ks, Ghorbani:2011xh, Ghorbani:2013xga}.} Here $Z_N^{\mbox{\scriptsize Nek}}(\epsilon_1,\epsilon_2)$ does not depend on the K\"ahler modulus $\zeta$ of the ADHM moduli space, whereas the stringy instanton partition function depends on $\zeta$. In Section \ref{sec:st_inst} we review the results in \cite{Bonelli:2013rja}, and discuss the formal structures of the stringy $U(N)$ instanton partition function.

The stringy instanton partition function has the classical stringy corrections and quantum stringy ($\alpha'$) corrections. In the anti-self-dual Omega background $\epsilon_1=-\epsilon_2=\hbar$, the stringy instanton partition function does not have the quantum stringy corrections, and only have the classical stringy corrections \cite{Bonelli:2013rja}. In Section \ref{sec:class_st_inst}, we study the classical part $Z_N^{\mbox{\scriptsize cSI}}(\epsilon_1,\epsilon_2;\zeta)$ of the stringy $U(N)$ instanton partition function (see (\ref{sinst_classic}) for the $k$-instanton sector), and show that $Z_N^{\mbox{\scriptsize cSI}}(\epsilon_1,\epsilon_2;\zeta)$ is reduced from a four dimensional limit of the K-theoretic (``five dimensional ${\IR}^4\times S^1$'') $U(N)$ instanton partition function $Z_{N,m}^{\mbox{\scriptsize K-Nek}}(\epsilon_1,\epsilon_2)$ with five dimensional Chern-Simons coefficient $m\in {\IZ}$ \cite{Tachikawa:2004ur, Gottsche:2006bm}. By geometric engineering, the classical stringy instanton partition function $Z_N^{\mbox{\scriptsize cSI}}(\epsilon_1,\epsilon_2;\zeta)$ is also reduced from the refined topological A-model partition function $Z_{N,m}^{\mbox{\scriptsize refA}}(\epsilon_1,\epsilon_2)$ on a family of local toric Calabi-Yau threefolds with resolved $A_{N-1}$ singularity labeled by $m\in {\IZ}$ described in Figure \ref{ladder} (Section \ref{subsec:rel_top_st}). Then we obtain a relation
$$
Z_{N,m}^{\mbox{\scriptsize refA}}(\epsilon_1,\epsilon_2)\sim Z_{N,m}^{\mbox{\scriptsize K-Nek}}(\epsilon_1,\epsilon_2) \ \stackrel{\mbox{\scriptsize 4d limit}}{\longrightarrow}\ Z_N^{\mbox{\scriptsize cSI}}(\epsilon_1,\epsilon_2;\zeta) \ \stackrel{r\to 0}{\longrightarrow}\ Z_N^{\mbox{\scriptsize Nek}}(\epsilon_1,\epsilon_2).
$$
Here the four dimensional (4d) limit is given by
$$
\beta \to 0, \ \ m\to \infty\ \ \mbox{with fixed}\ \ \beta m \sim \zeta r,
$$
where $\beta$ is the radius of the five dimensional circle $S^1$, and $r$ is the radius of the worldsheet $S^2$. The limit $r\to 0$ corresponds to the degenerate limit of the $S^2$. In Section \ref{subsec:brane_geom}, we give a physical explanation of the above relation by revisiting string dualities discussed in \cite{Karch:1998yv}. Then this relation claims geometric engineering of the instantons with the classical stringy corrections, and shows that the K\"ahler modulus $\zeta$ of the ADHM moduli space is obtained from the strong coupling limit of the five dimensional Chern-Simons term, and quantization of $\zeta$ can be interpreted as a five dimensional Chern-Simons coefficient $m\in{\IZ}$.

It was conjectured in \cite{Jockers:2012dk} and proved in \cite{Gomis:2012wy} that the ${\cal N}=(2,2)$ GLSM partition function $Z_X^{\mbox{\scriptsize GLSM}}$ on $S^2$ which flows in the IR fixed point to an ${\cal N}=(2,2)$ NLSM on $S^2$ whose target space is a Calabi-Yau geometry $X$ gives the quantum-corrected K\"ahler potential $K_X$ on the K\"ahler moduli space of $X$:
$$
e^{-K_X}=Z_X^{\mbox{\scriptsize GLSM}}.
$$
Therefore the stringy $U(N)$ instanton partition function gives the K\"ahler potential for the ADHM moduli space. Note that the ADHM moduli space has a hyper-K\"ahler structure \cite{NakajimaLec}, and thus satisfies the Calabi-Yau condition. In \cite{Bonelli:2013rja}, it was proposed that the quantum stringy corrections gives us the Givental's ${\cal I}$-function \cite{Givental} (see also \cite{CoatesGivental}) of the ADHM moduli space. In Section \ref{sec:quant_st_inst}, for $U(1)$, $U(2)$, and $U(3)$ we study the full stringy instanton partition functions with the quantum stringy corrections. For each $N$, we find a universal structure of the quantum stringy corrections for arbitrary instanton charge $k$. For $U(1)$, as discussed in \cite{Bonelli:2013rja}, we confirm agreements with quantum correlators in the $T$-equivariant cohomology ring $H_T^*\big(\mbox{Hilb}_k({\IC}^2),{\IQ}\big)$ (with the equivariant parameters $\epsilon_{1,2}$) of the Hilbert scheme $\mbox{Hilb}_k({\IC}^2)$ of points on ${\IC}^2$. For $U(2)$ and $U(3)$, using the formulation of Maulik and Okounkov \cite{Maulik:2012wi}, we also check agreements with the quantum parts of quantum correlators in the equivariant cohomology of the ADHM moduli space.
In Section \ref{sec:conclusion} we give our conclusions and discuss some future directions. Appendix \ref{app:multi_gamma} is a note on the multiple gamma function which appears in this paper. In Appendix \ref{app:st_inst_hurwitz}, we discuss relations between the stringy $U(1)$ instanton partition function, the simple Hurwitz theory, and the topological A-model on a local toric Calabi-Yau threefold. Here a relation between the latter two theories was previously discussed in \cite{Caporaso:2006gk}. In Appendix \ref{app:equiv_hilb}, we review the Fock space description of $H_T^*\big(\mbox{Hilb}_k({\IC}^2),{\IQ}\big)$ \cite{Grojnowski, NakHilb, NakajimaLec, OkounkovPandharipande}, and compute some equivariant correlators for comparing with the stringy $U(1)$ instanton partition function. In Appendix \ref{app:equiv_adhm}, we review and discuss the 
higher rank generalization of the formulation in Appendix \ref{app:equiv_hilb} \cite{Bar, Maulik:2012wi}.
In Appendix \ref{app:exact_kahler}, we summarize the formulas of the exact K\"ahler potentials on quantum K\"ahler moduli spaces of Calabi-Yau threefolds (e.g. \cite{Jockers:2012dk}) and fourfolds (conjectured in \cite{Honma:2013hma}).

\section{Stringy $U(N)$ instanton partition function}\label{sec:st_inst}

In this section, we review the stringy $U(N)$ instanton partition function given in \cite{Bonelli:2013rja}, and discuss its formal structures.

\subsection{${\cal N}=(2,2)$ GLSM on $S^2$ for ADHM moduli space}\label{subsec:n22_glsm}

Let us consider type IIB strings on ${\IC}^2\times T^*{\IP}^1\times {\IC}$, and introduce D1-D5 brane system consisting of $N$ D5-branes on ${\IC}^2\times{\IP}^1$ and $k$ D1-branes wrapping around the ${\IP}^1$. By embedding the $U(1)$ spin connection on ${\IP}^1$ into the $SO(4)$ R-symmetry in the world volume theory of the $N$ D5-branes,\footnote{This embedding breaks the R-symmetry as $SO(4)\cong SU(2)\times SU(2)$ to $U(1)\times SU(2)$.} and by compactifying this theory on ${\IC}^2$ the four dimensional ${\cal N}=2$ $U(N)$ supersymmetric gauge theory is obtained \cite{Bershadsky:1995qy} (see also e.g. \cite{DiVecchia:2002ks}). Here the $k$ D1-branes describe the instantons of charge $k$ in the four dimensional gauge theory.\footnote{This theory has the KK corrections by the compactification of the ${\IP}^1$, and these corrections are given as stringy corrections. By the degeneration of ${\IP}^1$, the D1-D5 brane system becomes the fractional D$(-1)$-D3 brane system at the ${\IC}^2/{\IZ}_2$ orbifold singularity \cite{Douglas:1996sw, Douglas:1996xg, Diaconescu:1997br} which describes the four dimensional ${\cal N}=2$ $U(N)$ supersymmetric gauge theory without the KK corrections.} As the world volume theory on the $k$ D1-branes, these instantons are described by an ${\cal N}=(2,2)$ NLSM on ${\IP}^1$ whose target space is the ADHM (framed $k$-instanton) moduli space\footnote{The ADHM moduli space has a hyper-K\"ahler structure and the complex dimension is given by $\dim_{{\IC}}{\cal M}_{k,N}=2kN$ \cite{NakajimaLec}.}
\begin{equation}
{\cal M}_{k,N}=\{(B_1,B_2,I,J) | [B_1,B_2]+IJ=0,\ [B_1,B_1^{\dagger}]+[B_2,B_2^{\dagger}]+II^{\dagger}-J^{\dagger}J=\zeta \mathbf{1}_{k\times k}\}/U(k).
\label{ADHM_moduli}
\end{equation}
Here $B_{1,2}$: ${\IC}^k \to {\IC}^k$, $I$: ${\IC}^N \to {\IC}^k$, $J$: ${\IC}^k \to {\IC}^N$, and $\zeta>0$ defines the K\"ahler modulus. The gauge transformation is given by $(B_1,B_2,I,J)\ \mapsto\ (R^{-1}B_1R,R^{-1}B_2R,R^{-1}I,JR)$, where $R\in U(k)$. This NLSM is obtained in the IR fixed point of an ${\cal N}=(2,2)$ $U(k)$ GLSM on $S^2\cong{\IP}^1$ with the matter content described in Table \ref{ADHMmat} \cite{Bonelli:2013rja}. The twisted masses $\epsilon_{1,2}\in{\IR}$ give the generators of $T^2=U(1)^2$ which rotates the ${\IC}^2$ in ${\IC}^2\times{\IP}^1$, and these masses induce the Omega background. The twisted masses $a_{\alpha}\in{\IR}$ give the generators of the Cartan subalgebra of $U(N)$. These chiral fields interact each other through a superpotential $W={\rm Tr}_k\chi([B_1,B_2]+IJ)$ whose total $U(1)_V$ R-charge is two. The Fayet-Iliopoulos (FI) parameter for central $U(1)\subset U(k)$ gives the K\"ahler modulus $\zeta$ in the ADHM moduli space (\ref{ADHM_moduli}).
\begin{table}[t]
\begin{center}
\begin{tabular}{c|c|c|c}
\hline
Field & $U(k)$ & Twisted mass& $U(1)_V$ \\ \hline
$\chi$ & ${\bf adj.}$ & $\epsilon$ & $2-2\mathfrak{q}$ \\
$B_{1,2}$ & ${\bf adj.}$ & $-\epsilon_{1,2}$ & $\mathfrak{q}$ \\
$I_{\alpha}$ & $\mathbf{k}$ & $-a_{\alpha}$ & $\mathfrak{q}+\mathfrak{p}$ \\
$J_{\alpha}$ & $\mathbf{\overline{k}}$ & $a_{\alpha}-\epsilon$ & $\mathfrak{q}-\mathfrak{p}$ \\\hline
\end{tabular}
\caption{Matter content of the GLSM for the ADHM moduli space ${\cal M}_{k,N}$. Here $\epsilon=\epsilon_1+\epsilon_2$ and $\alpha=1, \ldots, N$. By restricting the $U(1)_V$ R-charges to be non-negative, these are constrained as $0<\mathfrak{p}<\mathfrak{q}<1$.}
\label{ADHMmat}
\end{center}
\end{table}

By using the formula of the GLSM partition function on $S^2$ obtained by the supersymmetric localization \cite{Benini:2012ui,Doroud:2012xw}, after taking the limit $\mathfrak{p},\mathfrak{q} \to 0^{+}$ due to the non-compactness of the ${\cal M}_{k,N}$ \cite{Park:2012nn}, one obtains the stringy $U(N)$ $k$-instanton partition function \cite{Bonelli:2013rja}:
\begin{equation}
Z_{k,N}(\epsilon_1,\epsilon_2,\vec{a},z)=\frac{1}{k!}\sum_{\overrightarrow{m}\in {\IZ}^k}\int_{{\IR}^k}\bigg[\prod_{a=1}^k\frac{d\sigma_a}{2\pi}z^{i\sigma_a-\frac{m_a}{2}}\overline{z}^{i\sigma_a+\frac{m_a}{2}}\bigg]
\bigg[\prod_{a<b}^k\Big(\frac{m_{ab}^2}{4}+\sigma_{ab}^2\Big)\bigg]Z_{IJ}Z_{\mbox{\scriptsize adj}},
\end{equation}
where $z=e^{-2\pi\zeta+i\theta}$ with the theta angle $\theta$, $m_{ab}=m_a-m_b$, and $\sigma_{ab}=\sigma_a-\sigma_b$. Here
\begin{equation}
Z_{IJ}=\prod_{a=1}^k\prod_{\alpha=1}^N\frac{\Gamma(0^{+}-i\sigma_a+ira_{\alpha}-\frac12m_a)}{\Gamma(1+i\sigma_a-ira_{\alpha}-\frac12m_a)}\frac{\Gamma(0^{+}+i\sigma_a-ir(a_{\alpha}-\epsilon)+\frac12m_a)}{\Gamma(1-i\sigma_a+ir(a_{\alpha}-\epsilon)+\frac12m_a)}
\end{equation}
is the one loop determinant of the chiral multiplets including the chiral fields $I_{\alpha}$, $J_{\alpha}$, and
\begin{equation}
Z_{\mbox{\scriptsize adj}}=\prod_{a,b=1}^k\frac{\Gamma(1-i\sigma_{ab}-ir\epsilon-\frac12m_{ab})}{\Gamma(i\sigma_{ab}+ir\epsilon-\frac12m_{ab})}\frac{\Gamma(-i\sigma_{ab}+ir\epsilon_1-\frac12m_{ab})}{\Gamma(1+i\sigma_{ab}-ir\epsilon_1-\frac12m_{ab})}\frac{\Gamma(-i\sigma_{ab}+ir\epsilon_2-\frac12m_{ab})}{\Gamma(1+i\sigma_{ab}-ir\epsilon_2-\frac12m_{ab})}
\end{equation}
is the one loop determinant of the chiral multiplets including the chiral fields $\chi$,  $B_{1,2}$, where $r$ is the radius of $S^2$, and $\epsilon=\epsilon_1+\epsilon_2$.

In the large radius phase $\zeta \gg 0$, the above partition function can be written as \cite{Bonelli:2013rja}
\begin{equation}
Z_{k,N}(\epsilon_1,\epsilon_2,\vec{a},z)=\frac{1}{k!}\oint\bigg[\prod_{a=1}^k\frac{d(r\lambda_a)}{2\pi i}(z\overline{z})^{-r\lambda_a}\bigg]Z_{\mbox{\scriptsize L}}|Z_{\mbox{\scriptsize V}}(z)|^2,
\label{sinst}
\end{equation}
where
\begin{align}
&
Z_{\mbox{\scriptsize L}}=\bigg(\frac{\Gamma(1-ir\epsilon)\Gamma(ir\epsilon_1)\Gamma(ir\epsilon_2)}{\Gamma(ir\epsilon)\Gamma(1-ir\epsilon_1)\Gamma(1-ir\epsilon_2)}\bigg)^k\prod_{a=1}^k\prod_{\alpha=1}^N\frac{\Gamma(r\lambda_a+ira_{\alpha})\Gamma(0^{+}-r\lambda_a-ir(a_{\alpha}-\epsilon))}{\Gamma(1-r\lambda_a-ira_{\alpha})\Gamma(1+r\lambda_a+ir(a_{\alpha}-\epsilon))}\nonumber\\
&\hspace{3em}\times
\prod_{\begin{subarray}{c}a,b=1\\(a\neq b)\end{subarray}}^kr\lambda_{ab}\frac{\Gamma(1+r\lambda_{ab}-ir\epsilon)\Gamma(r\lambda_{ab}+ir\epsilon_1)\Gamma(r\lambda_{ab}+ir\epsilon_2)}{\Gamma(-r\lambda_{ab}+ir\epsilon)\Gamma(1-r\lambda_{ab}-ir\epsilon_1)\Gamma(1-r\lambda_{ab}-ir\epsilon_2)},
\label{sinst_loop}
\\
&
Z_{\mbox{\scriptsize V}}(z)=\sum_{d_1,\ldots,d_k=0}^{\infty}z^{\sum_{a=1}^kd_a}\prod_{a=1}^k\prod_{\alpha=1}^N\frac{(-r\lambda_a-ir(a_{\alpha}-\epsilon))_{d_a}}{(1-r\lambda_a-ira_{\alpha})_{d_a}}\nonumber\\
&\hspace{3em}\times
\prod_{1\le a<b\le k}\frac{d_{ba}+r\lambda_{ab}}{r\lambda_{ab}}\frac{(1+r\lambda_{ab}-ir\epsilon)_{d_{ba}}(r\lambda_{ab}+ir\epsilon_1)_{d_{ba}}(r\lambda_{ab}+ir\epsilon_2)_{d_{ba}}}{(r\lambda_{ab}+ir\epsilon)_{d_{ba}}(1+r\lambda_{ab}-ir\epsilon_1)_{d_{ba}}(1+r\lambda_{ab}-ir\epsilon_2)_{d_{ba}}},
\label{sinst_vor}
\end{align}
$\lambda_{ab}=\lambda_a-\lambda_b$, $d_{ab}=d_a-d_b$, and $(x)_d=\frac{\Gamma(x+d)}{\Gamma(x)}$ is the Pochhammer symbol. Here by shifting the theta angle $\theta$, we have scaled $z$ as $z \to (-1)^{N+k-1}z$ for simplicity. The contours in (\ref{sinst}) are enclosing the imaginary axes counterclockwise, and the simple poles which give residues only come from the numerators in (\ref{sinst_loop}) given by $\lambda_a=-ia_{\alpha}$, $\lambda_{ab}=-i\epsilon_1$, and $\lambda_{ab}=-i\epsilon_2$. As the result the simple poles are labeled by an $N$-tuple of Young diagrams $\vec{\mu}=(\mu_1,\ldots,\mu_N)$ with $k=|\vec{\mu}|=\sum_{\alpha=1}^N|\mu_{\alpha}|$, where $|\mu_{\alpha}|$ is the number of boxes of the Young diagram $\mu_{\alpha}$.

As an example, in the $N=2$ and $k=3$ case, we see that the simple poles are classified by ten types of two-tuple of Young diagrams $\vec{\mu}_{1,\ldots,10}$ as
\begin{align}
&
\vec{\mu}_1=(\tableau{1 1 1},\bullet) :\ \lambda_1=-ia_1,\ \lambda_2=-ia_1-i\epsilon_1,\ \lambda_3=-ia_1-2i\epsilon_1,\nonumber\\
&
\vec{\mu}_2=(\tableau{2 1},\bullet) :\ \lambda_1=-ia_1,\ \lambda_2=-ia_1-i\epsilon_1,\ \lambda_3=-ia_1-i\epsilon_2,\nonumber\\
&
\vec{\mu}_3=(\tableau{1 1},\tableau{1}) :\ \lambda_1=-ia_1,\ \lambda_2=-ia_1-i\epsilon_1,\ \lambda_3=-ia_2,
\end{align}
$\vec{\mu}_4=(\tableau{3},\bullet)$ obtained by $\epsilon_1\leftrightarrow\epsilon_2$ in $\vec{\mu}_1$, $\vec{\mu}_5=(\tableau{2},\tableau{1})$ obtained by $\epsilon_1\leftrightarrow\epsilon_2$ in $\vec{\mu}_3$, and $\vec{\mu}_{6,\ldots,10}=(\bullet,\tableau{1 1 1})$, $(\bullet,\tableau{2 1})$, $(\tableau{1},\tableau{1 1})$, $(\bullet, \tableau{3})$, $(\tableau{1},\tableau{2})$ obtained by $a_1\leftrightarrow a_2$ in $\vec{\mu}_{1,\ldots,5}$.

\begin{figure}[t]
 \begin{center}
  \includegraphics[width=55mm]{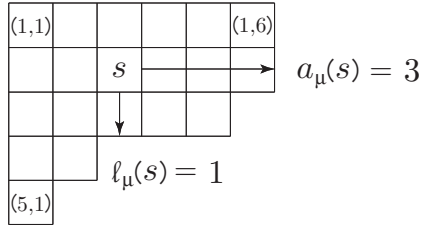}
 \end{center}
 \caption{An example of Young diagram $\mu$ with $|\mu|=20$. Let $s=(i,j)\in \mu$ be the label of the box at the $i$-th low and the $j$-th column of $\mu$. At $s=(2,3)$, $a_{\mu}(s)=3$ and ${\ell}_{\mu}(s)=1$.}
\label{YoungNek}
\end{figure}

Let us consider the degenerate limit $r\to 0$ of $S^2$. Under this limit $(z\overline{z})^{-r\lambda_a}=1+{\cal O}(r)$,
\begin{equation}
Z_{\mbox{\scriptsize V}}(z)=1+{\cal O}(r^N),
\label{ZV_behavior}
\end{equation}
and using $\Gamma(r)=r^{-1}+{\cal O}(r^0)$ to estimate $Z_{\mbox{\scriptsize L}}$, one finds that the stringy $U(N)$ $k$-instanton partition function (\ref{sinst}) yields the $U(N)$ $k$-instanton (Nekrasov) partition function $Z_{k,N}^{\mbox{\scriptsize Nek}}$ \cite{Bonelli:2013rja}:
\begin{equation}
Z_{k,N}(\epsilon_1,\epsilon_2,\vec{a},z)=\frac{1}{(ir)^{2kN}}Z_{k,N}^{\mbox{\scriptsize Nek}}(\epsilon_1,\epsilon_2,\vec{a})+\mbox{stringy corrections},
\label{sinst_expNek}
\end{equation}
where
\begin{align}
Z_{k,N}^{\mbox{\scriptsize Nek}}(\epsilon_1,\epsilon_2,\vec{a})=\frac{\epsilon^k}{k!(2\pi i\epsilon_1\epsilon_2)^k}\oint\prod_{a=1}^{k}&
\frac{d\sigma_a}{\prod_{\alpha=1}^N(\sigma_a+a_{\alpha})(i0^{+}+\epsilon-(\sigma_a+a_{\alpha}))}\nonumber\\
&\times
\prod_{1\le a<b\le k}\frac{\sigma_{ab}^2(\sigma_{ab}^2-\epsilon^2)}{(\sigma_{ab}^2-\epsilon_1^2)(\sigma_{ab}^2-\epsilon_2^2)},
\label{Nekrasov_integ}
\end{align}
and the contours are enclosing the real axes counterclockwise \cite{Moore:1997dj, Moore:1998et, Nekrasov:2002qd}. The exponent of the leading behavior $r^{-2kN}$ coincides with $\dim_{{\IC}}{\cal M}_{k,N}=2kN$. It is useful to express the Nekrasov partition function by using $N$-tuple of Young diagrams $\vec{\mu}=(\mu_1,\ldots,\mu_N)$ as \cite{Nekrasov:2002qd, Flume:2002az, Bruzzo:2002xf, Nakajima:2003pg}
\begin{equation}
Z_{k,N}^{\mbox{\scriptsize Nek}}(\epsilon_1,\epsilon_2,\vec{a})=\sum_{|\vec{\mu}|=k}\frac{1}{\prod_{\alpha,\beta=1}^Nn_{\alpha\beta}^{\vec{\mu}}(\epsilon_1,\epsilon_2,\vec{a})},
\label{Nekrasov_in}
\end{equation}
where
\begin{equation}
n_{\alpha\beta}^{\vec{\mu}}(\epsilon_1,\epsilon_2,\vec{a})=
\prod_{s\in \mu_{\alpha}}\left(a_{\alpha\beta}-\epsilon_1\ell_{\mu_{\beta}}(s)
+\epsilon_2(a_{\mu_{\alpha}}(s)+1)\right)
\prod_{t\in \mu_{\beta}}
\left(a_{\alpha\beta}+
\epsilon_1(\ell_{\mu_{\alpha}}(t)+1)-\epsilon_2a_{\mu_{\beta}}(t)\right).
\end{equation}
Here $a_{\alpha\beta}=a_{\alpha}-a_{\beta}$, and $a_{\mu}(s)$ and $\ell_{\mu}(s)$ are the arm- and leg-length, respectively. In Figure \ref{YoungNek} we describe an example of Young diagram $\mu$ with $|\mu|=20$. The expansion (\ref{sinst_expNek}) contains the stringy corrections ${\cal O}(r^{-2kN+1})$ to the Nekrasov partition function, and the main subject of this paper is to study these corrections.

\subsection{Stringy corrections}\label{subsec:st_corr}

As discussed in \cite{Bonelli:2013rja}, $Z_{\mbox{\scriptsize L}}$ in (\ref{sinst_loop}) contains the perturbative $\alpha'$ corrections. In the supersymmetric localization \cite{Benini:2012ui, Doroud:2012xw}, these corrections depends on the regularization scheme of the infinite products in the one loop determinant of the chiral multiplets. In \cite{Benini:2012ui, Doroud:2012xw} this ambiguity was fixed by the zeta function regularization $\left[\prod_{n=0}^{\infty}(x+n)\right]_{\mbox{\scriptsize reg}}=\frac{\sqrt{2\pi}}{\Gamma(x)}$ (see Appendix \ref{subapp:zeta}). By
\begin{equation}
\frac{\Gamma(-x)}{\Gamma(x)}=-\exp\Big(2\gamma x+2\sum_{s=1}^{\infty}\frac{\zeta(2s+1)}{2s+1}x^{2s+1}\Big),
\label{Gamma_zeta}
\end{equation}
we see that the perturbative $\alpha'$ corrections corrections in (\ref{sinst_expNek}) contain the Euler constant $\gamma$ and the simple zeta values $\zeta(s)=\sum_{n=1}^{\infty}\frac{1}{n^s}$. On the other hand, $Z_{\mbox{\scriptsize V}}(z)$ in (\ref{sinst_vor}) contains the non-perturbative $\alpha'$ corrections and does not depend on the regularization scheme. Especially it was proposed that $Z_{\mbox{\scriptsize V}}(z)$ gives the Givental's ${\cal I}$-function \cite{Givental, CoatesGivental} of the ADHM moduli space ${\cal M}_{k,N}$ \cite{Bonelli:2013rja, Bonelli:2013mma}:
\begin{equation}
{\cal I}(z)=Z_{\mbox{\scriptsize V}}(z).
\label{Iz_ZV}
\end{equation}
Here $\lambda_1,\ldots,\lambda_k$ are identified with the Chern roots of the tautological bundle on ${\cal M}_{k,N}$ (for $N=1$, see \cite{Fontanine}). The inverse radius $r^{-1}$ of the two sphere $S^2$ is identified with the equivariant parameter which gives the generator of $S^1$ acting on the $S^2$. By the equivariant mirror map, the ${\cal I}$-function is related with the small ${\cal J}$-function ${\cal J}(z)$. Then by expanding the ${\cal J}$-function around $r=0$, one can obtain the $T^{N+2}$-equivariant Gromov-Witten invariants.\footnote{The $N+2$ dimensional torus $T^{N+2}=U(1)^N\times U(1)^2$ acts on ${\cal M}_{k,N}$ as $U(1)^N$: $(B_1, B_2, I, J)\ \mapsto\ (B_1, B_2, IT_a^{-1}, T_aJ)$ where $T_a=\mbox{diag}(e^{ia_1},\ldots,e^{ia_N})$, and $U(1)^2$: $(B_1, B_2, I, J)\ \mapsto\ (T_{\epsilon_1}B_1, T_{\epsilon_2}B_2, I, T_{\epsilon_1}T_{\epsilon_2}J)$ where $T_{\epsilon_{1,2}}=e^{i\epsilon_{1,2}}$. This action introduces the equivariant parameters $a_{\alpha}$, $\alpha=1,\ldots,N$ and $\epsilon_{1,2}$ corresponding to the twisted masses in the GLSM as described in Table \ref{ADHMmat}.}
It is known that the coefficient of $r^1$ in the expansion around $r=0$ of the ${\cal I}$-function gives the equivariant mirror map. As discussed in \cite{Bonelli:2013rja}, by (\ref{Iz_ZV}) and the behavior (\ref{ZV_behavior}) of $Z_{\mbox{\scriptsize V}}(z)$, one finds that
the equivariant mirror maps for the $N\ge2$ cases are trivial: ${\cal J}(z)={\cal I}(z)=Z_{\mbox{\scriptsize V}}(z)$, but for the $N=1$ case the equivariant mirror map is needed as \cite{Fontanine}
\begin{equation}
{\cal J}(z)=(1-z)^{ikr\epsilon}{\cal I}(z)=(1-z)^{ikr\epsilon}Z_{\mbox{\scriptsize V}}(z).
\label{equiv_mirror1}
\end{equation}

In (\ref{sinst}), by taking the leading terms of $Z_{\mbox{\scriptsize L}} \sim {\cal O}(r^{-2kN})$ and $Z_{\mbox{\scriptsize V}}(z) \sim 1$ around $r=0$, we obtain the stringy instanton partition function without the quantum stringy ($\alpha'$) corrections:
\begin{equation}
Z_{k,N}^{\mbox{\scriptsize classic}}(\epsilon_1,\epsilon_2,\vec{a},z)=\frac{1}{(ir)^{2kN}}\sum_{|\vec{\mu}|=k}\frac{(z\overline{z})^{ir\sum_{\alpha=1}^N\left(a_{\alpha}|\mu_{\alpha}|+\epsilon_1n(\mu_{\alpha})+\epsilon_2n(\mu_{\alpha}^t)\right)}}{\prod_{\alpha,\beta=1}^Nn_{\alpha\beta}^{\vec{\mu}}(\epsilon_1,\epsilon_2,\vec{a})},
\label{sinst_classic}
\end{equation}
where $n(\mu)=\sum_{s\in\mu}\ell_{\mu}(s)$ and $n(\mu^t)=\sum_{s\in\mu}a_{\mu}(s)$.
Especially in the anti-self-dual case $\epsilon_1=-\epsilon_2=\hbar$, one finds that the stringy instanton partition function does not have the $\alpha'$ corrections \cite{Bonelli:2013rja} (see also \cite{Maulik:2012wi}), and thus
\begin{equation}
Z_{k,N}(\hbar,-\hbar,\vec{a},z)=Z_{k,N}^{\mbox{\scriptsize classic}}(\hbar,-\hbar,\vec{a},z).
\label{sinst_ASD}
\end{equation}
In Section \ref{sec:class_st_inst} and Section \ref{sec:quant_st_inst} we study the classical and quantum stringy corrections, respectively.

\section{Classical stringy corrections}\label{sec:class_st_inst}

In this section we study the classical stringy corrections. Let us define the generating function of the classical stringy $U(N)$ instanton partition functions (\ref{sinst_classic}) by
\begin{equation}
Z_N^{\mbox{\scriptsize cSI}}(\epsilon_1,\epsilon_2,\vec{a},z,\Lambda)=1+\sum_{k=1}^{\infty}(r\Lambda)^{2kN}Z_{k,N}^{\mbox{\scriptsize classic}}(\epsilon_1,\epsilon_2,\vec{a},z),
\label{sinst_gen_class}
\end{equation}
where $\Lambda$ is the dynamical scale in four dimensional gauge theory. This partition function gives the instanton (``non-perturbative'' in the gauge theoretic sense) partition function without the $\alpha'$ corrections of the $N$ D5-world volume theory on ${\IC}^2\times{\IP}^1$ that was dimensionally reduced to the ${\IC}^2$. In \cite{Bonelli:2013rja}, the ``perturbative'' (in the gauge theoretic sense) partition function
\begin{equation}
Z_N^{\mbox{\scriptsize D5}}(\epsilon_1,\epsilon_2,\vec{a})=Z_N^{\mbox{\scriptsize cD5}}(\epsilon_1,\epsilon_2,\vec{a})\times Z_N^{\mbox{\scriptsize qD5}}(\epsilon_1,\epsilon_2,\vec{a})
\end{equation}
was also computed. Here
\begin{equation}
Z_N^{\mbox{\scriptsize cD5}}(\epsilon_1,\epsilon_2,\vec{a})=\prod_{\begin{subarray}{c}\alpha,\beta=1\\(\alpha\neq \beta)\end{subarray}}^N
\Gamma_2(a_{\alpha\beta}|\epsilon_1,\epsilon_2)
\label{class_D5}
\end{equation}
gives the well-known perturbative partition function of the four dimensional ${\cal N}=2$ $U(N)$ gauge theory on the Omega background \cite{Nekrasov:2003rj, Nakajima:2003uh} (see also Appendix \ref{subapp:pert_4d}). The quantum stringy part
\begin{equation}
Z_N^{\mbox{\scriptsize qD5}}(\epsilon_1,\epsilon_2,\vec{a})=\prod_{\begin{subarray}{c}\alpha,\beta=1\\(\alpha\neq \beta)\end{subarray}}^N
\frac{\Gamma_3(a_{\alpha\beta}|\epsilon_1,\epsilon_2,\frac{1}{ir})}{\Gamma_3(a_{\alpha\beta}|\epsilon_1,\epsilon_2,-\frac{1}{ir})}
\end{equation}
contains the perturbative $\alpha'$ corrections, where $\Gamma_r(z|\omega_1,\ldots,\omega_r)$ is the multiple gamma function (see Appendix \ref{app:multi_gamma}), and one finds $Z_N^{\mbox{\scriptsize qD5}}(\hbar,-\hbar,\vec{a})=1$. Then in the anti-self-dual case $\epsilon_1=-\epsilon_2=\hbar$, there are no $\alpha'$ corrections as in the instanton partition function (\ref{sinst_ASD}),
\begin{equation}
Z_N^{\mbox{\scriptsize D5}}(\hbar,-\hbar,\vec{a})=Z_N^{\mbox{\scriptsize cD5}}(\hbar,-\hbar,\vec{a}).
\label{D5_ASD}
\end{equation}

In the following we show that the classical stringy partition functions (\ref{sinst_gen_class}) (and (\ref{class_D5})) are reduced from a four dimensional limit of the K-theoretic Nekrasov partition function with five dimensional Chern-Simons term. We also find that the instantons with the classical stringy corrections can be embedded into (refined) topological string theory.

\subsection{Relation with K-theoretic instanton partition function}\label{subsec:rel_k_inst}

The K-theoretic $U(N)$ instanton (Nekrasov) partition function with five dimensional Chern-Simons term $\int{\rm Tr}A\wedge F\wedge F$ \cite{Intriligator:1997pq} is given by \cite{Tachikawa:2004ur, Gottsche:2006bm}
\begin{equation}
Z_{N,m}^{\mbox{\scriptsize K-Nek}}(\epsilon_1,\epsilon_2,\vec{a},\Lambda)=\sum_{\vec{\mu}}\frac{\Big(e^{-\frac12(N-m)\epsilon}\Lambda^{2N}\Big)^{|\vec{\mu}|}}{\prod_{\alpha,\beta=1}^NN_{\alpha\beta}^{\vec{\mu}}(\epsilon_1,\epsilon_2,\vec{a})}\cdot e^{m\sum_{\alpha=1}^N\left(a_{\alpha}|\mu_{\alpha}|+\epsilon_1n(\mu_{\alpha})
+\epsilon_2n(\mu_{\alpha}^t)\right)},
\label{K_Nek_CS}
\end{equation}
where $m\in{\IZ}$ is the Chern-Simons coefficient, and
\begin{equation}
N_{\alpha\beta}^{\vec{\mu}}(\epsilon_1,\epsilon_2,\vec{a})=\prod_{s\in \mu_{\alpha}}
\left(1-Q_{\alpha\beta}t^{-\ell_{\mu_{\beta}}(s)}q^{-a_{\mu_{\alpha}}(s)-1}\right)
\prod_{t\in \mu_{\beta}}
\left(1-Q_{\alpha\beta}t^{\ell_{\mu_{\alpha}}(t)+1}q^{a_{\mu_{\beta}}(t)}\right),
\end{equation}
$t=e^{-\epsilon_1}$, $q=e^{\epsilon_2}$, $Q_{\alpha\beta}=e^{-a_{\alpha\beta}}$. Let $\beta$ be the radius of the five dimensional circle. After scaling $a_{\alpha} \to i\beta a_{\alpha}, \epsilon_{1,2} \to i\beta \epsilon_{1,2}$, $\Lambda \to i e^{-\frac{im}{4N}\beta\epsilon}\beta\Lambda$, by taking a four dimensional limit
\begin{equation}
\beta \to 0, \ \ m\to \infty\ \ \mbox{with fixed}\ \ \beta m=r\log z\overline{z}=-4\pi \zeta r,
\label{4d_lim}
\end{equation}
we see that the K-theoretic Nekrasov partition function yields the classical stringy instanton partition function (\ref{sinst_gen_class}):
\begin{equation}
Z_{N,m}^{\mbox{\scriptsize K-Nek}}(\epsilon_1,\epsilon_2,\vec{a},\Lambda)\ \longrightarrow \ Z_N^{\mbox{\scriptsize cSI}}(\epsilon_1,\epsilon_2,\vec{a},z,\Lambda).
\label{KNek_Stringy_rel}
\end{equation}
The four dimensional limit (\ref{4d_lim}) relates (the strong coupling limit of) the Chern-Simons coefficient $m$ to the K\"ahler modulus $\zeta$. The K-theoretic ``perturbative'' (in the gauge theoretic sense) partition function $Z_{N}^{\mbox{\scriptsize K-pert}}$ \cite{Nekrasov:2003rj, Nakajima:2005fg, Gottsche:2006bm} does not depend on the Chern-Simons term, and yields the perturbative partition function $Z_N^{\mbox{\scriptsize cD5}}$ in (\ref{class_D5}) without the $\alpha'$ corrections. Note that in the anti-self-dual case $\epsilon_1=-\epsilon_2=\hbar$, as in (\ref{sinst_ASD}) and (\ref{D5_ASD}) the stringy partition functions do not receive the $\alpha'$ corrections, and the limit (\ref{4d_lim}) of the K-theoretic perturbative/non-perturbative (instanton) partition functions completely coincide with the stringy partition functions.

\subsection{Relation with topological strings}\label{subsec:rel_top_st}

By geometric engineering \cite{Klemm:1996bj, Katz:1996fh, Katz:1997eq}, it is known that the K-theoretic $SU(N)$ Nekrasov partition function with a Chern-Simons coefficient $m$ coincides with the partition function $Z_{N,m}^{\mbox{\scriptsize refA}}$ of the refined topological A-model on a local toric Calabi-Yau threefold $X_{N,m}$ given by ALE space fibration of $A_{N-1}$ type over ${\IP}^1$ \cite{Iqbal:2007ii, Taki:2007dh, Awata:2008ed} (see \cite{Iqbal:2003ix, Iqbal:2003zz, Eguchi:2003sj, Hollowood:2003cv, Eguchi:2003it, Zhou:2003zp} for the unrefined case). Topological type of this $SU(N)$ geometry $X_{N,m}$ described in Figure \ref{ladder} is labeled by an integer $m$ which is identified with the Chern-Simons coefficient \cite{Tachikawa:2004ur}. The $SU(N)$ geometry $X_{N,m}$ has one modulus $T_b$ of the base ${\IP}^1$ and $N-1$ moduli $T_{f_{\alpha}}, \alpha=1,\ldots,N-1$ of the fiber consisting of $N-1$ resolved ${\IP}^1$'s. Here $T_b$ is identified with the dynamical scale $\Lambda$, and $T_{f_{\alpha}}$ are identified with $N-1$ independent Coulomb moduli $a_{\alpha}$ in the $SU(N)$ gauge theory \cite{Katz:1996fh}:
\begin{figure}[t]
 \begin{center}
  \includegraphics[width=120mm]{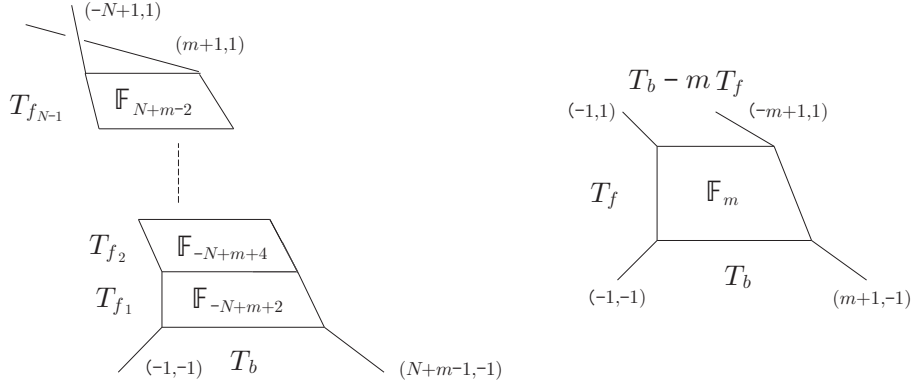}
 \end{center}
 \caption{The left figure describes the toric (web) diagram of the $SU(N)$ geometry $X_{N,m}$ labeled by $m\in{\IZ}$. It contains $N-1$ Hirzebruch surfaces $\{{\IF}_{-N+m+2\ell}\}_{\ell=1}^{N-1}$ as the compact divisors. The right figure describes the Hirzebruch surface ${\IF}_m$ with two K\"ahler moduli $T_f$ and $T_b$.}
\label{ladder}
\end{figure}
\begin{equation}
e^{-T_b}\sim (\beta\Lambda)^{2N},\ \ \ \
T_{f_{\alpha}}\sim \beta a_{\alpha\alpha+1}=\beta (a_{\alpha}-a_{\alpha+1}).
\end{equation}
Then by taking the limit (\ref{4d_lim}), we find that the partition function $Z_{N,m}^{\mbox{\scriptsize refA}}$ of the refined A-model on the family of the local toric Calabi-Yau threefolds $X_{N,m}$ labeled by $m$ yields the classical stringy partition function of the four dimensional $SU(N)$ gauge theory:
\begin{equation}
Z_{N,m}^{\mbox{\scriptsize refA}}(\epsilon_1,\epsilon_2,T_{f_{\alpha}},T_b)\ \longrightarrow \ Z_N^{\mbox{\scriptsize cD5}}(\epsilon_1,\epsilon_2,\vec{a})\times
Z_N^{\mbox{\scriptsize cSI}}(\epsilon_1,\epsilon_2,\vec{a},z,\Lambda)\Big|_{\scriptsize \sum_{\alpha=1}^Na_{\alpha}=0}.
\label{Top_Stringy_rel}
\end{equation}
The anti-self-dual case $\epsilon_1=-\epsilon_2=\hbar$ corresponds to the topological string (unrefined) limit, and $\hbar$ is identified with the topological string coupling constant $g_s$. Therefore the (unrefined) A-model partition function under the four dimensional limit (\ref{4d_lim}) completely coincides with the stringy partition function on the anti-self-dual Omega background.\footnote{A similar limit to (\ref{4d_lim}) was previously discussed in \cite{Caporaso:2006gk} to relate the simple Hurwitz theory with the topological A-model on the local toric Calabi-Yau threefold $X_{1,m}={\cal O}(m-2)\oplus{\cal O}(-m)\to {\IP}^1$. In Appendix \ref{app:st_inst_hurwitz}, we discuss and summarize such relations.}

\subsection{Brane construction and geometric engineering}\label{subsec:brane_geom}

In this section by revisiting a correspondence discussed in \cite{Karch:1998yv} between the world volume theory of $N$ D5-branes on ${\IC}^2\times$ $\{$vanishing ${\IP}^1\}$ and the geometrically engineered quantum field theory by local Calabi-Yau threefold with resolved $A_{N-1}$ singularity, we give a physical explanation of the relations (\ref{KNek_Stringy_rel}) and (\ref{Top_Stringy_rel}). As described in Section \ref{sec:st_inst}, 
the intersecting D1- and D5-branes in type IIB string theory on ${\IC}^2\times T^*{\IP}^1\times {\IC}$ lead to the instantons with the stringy corrections. In the D5-world volume theory, after the compactification of the ${\IP}^1$, to preserve eight supercharges in the four dimensions, the $U(1)$ spin connection on the ${\IP}^1$ needs to be embedded into the $SO(4)$ R-symmetry \cite{Bershadsky:1995qy}. This topological twist breaks the R-symmetry as $SO(4)\cong SU(2)\times SU(2)$ to $U(1)\times SU(2)$, and one obtains six dimensional ${\cal N}=(1,0)$ $U(N)$ supersymmetric gauge theory on ${\IC}^2\times {\IP}^1$ which leads to four dimensional ${\cal N}=2$ supersymmetric gauge theory.\vspace{0.5em}

\noindent\underline{\bf Type IIB theory $\to$ Type IIB theory:}\ \
Firstly, by taking the S-duality in type IIB string theory, the $N$ D5- and $k$ D1-branes are turned into NS5-branes and the fundamental strings wrapping around the compactified ${\IP}^1$.\vspace{0.5em}

\noindent\underline{\bf Type IIB theory $\to$ Type IIA theory:}\ \
As was shown in \cite{Ooguri:1995wj}, the $N$ NS5-branes in type IIB (IIA) string theory is equivalent to type IIA (IIB) string theory on an $A_{N-1}$ ALE space. By this duality, we obtain type IIA string theory compactified on a local Calabi-Yau threefold given by ALE space fibration of $A_{N-1}$ type over ${\IP}^1$, and then four dimensional ${\cal N}=2$ $SU(N)$ supersymmetric gauge theory is geometrically engineered \cite{Klemm:1996bj, Katz:1996fh, Katz:1997eq}. The low energy type IIA supergravity theory has the Chern-Simons term $\int B^{(2)}\wedge dC^{(3)}\wedge dC^{(3)}$, where $B^{(2)}$ is the NS-NS $B$ field and $C^{(3)}$ is the R-R 3-form field. From this Chern-Simons term, by integration on the local Calabi-Yau threefold, we obtain
\begin{equation}
\zeta_b g_{\alpha\beta\gamma} \int \phi_{\alpha}dA_{\beta} \wedge dA_{\gamma}.
\label{CS_int_B}
\end{equation}
Here $\phi_{\alpha}=\frac{1}{\zeta_b}\int_{{\cal C}_{\alpha}}B^{(2)}$ and $A_{\alpha}=\int_{{\cal C}_{\alpha}}C^{(3)}$ defined for the two cycles ${\cal C}_{\alpha}$, $\alpha=1,\ldots,N-1$ on the ALE space give scalar fields and gauge fields on ${\IC}^2$, respectively, and $g_{\alpha\beta\gamma}$ is given by a classical triple intersection number of divisors on the local Calabi-Yau threefold.\footnote{When the two cycles ${\cal C}_{\alpha}$ shrink to points, the $U(1)^{N-1}$ gauge symmetry is enhanced to $SU(N)$ \cite{Bershadsky:1995sp}.} It is known that the NS-NS $B$ field introduces a noncommutative parameter in the four dimensions, and it also introduces the FI parameter in the ADHM moduli space \cite{Aharony:1997an, Nekrasov:1998ss, Seiberg:1999vs}. Therefore we argue that the coefficient $\zeta_b g_{\alpha\beta\gamma}$ is identified with the FI parameter.

Let $X_N(\zeta)$ be the local Calabi-Yau threefold obtained from the toric Calabi-Yau threefold $X_{N,m}$ in Figure \ref{ladder} at the phase under the limit (\ref{4d_lim}).\footnote{The parameter $m$ in $X_{N,m}$ gives a classical triple intersection number.}
By considering this Calabi-Yau threefold $X_N(\zeta)$, as discussed above, we find an interaction term like (\ref{CS_int_B}).
As discussed in \cite{Tachikawa:2004ur}, the parameter $\zeta$ obtained by the scaling of the intersection number $m$ in $X_{N,m}$
is related with the coefficient of (\ref{CS_int_B}), and so
identified with the FI parameter. Then we see that the instanton partition function with the term (\ref{CS_int_B}) on the Omega background is given by the classical stringy instanton partition functions (\ref{sinst_classic}):
\begin{equation}
\int_{{\cal M}_{k,N}}e^{i\zeta \phi}\Big|_{\scriptsize \sum_{\alpha=1}^Na_{\alpha}=0}=Z_{k,N}^{\mbox{\scriptsize classic}}(\epsilon_1,\epsilon_2,\vec{a},z)\Big|_{\scriptsize \sum_{\alpha=1}^Na_{\alpha}=0},
\end{equation}
where $e^{i\zeta \phi}$ is an element of the $T=U(1)^N\times U(1)^2$-equivariant cohomology $H_T^*\big({\cal M}_{k,N}\big)$ \cite{Moore:1997dj}. Note that we also have the fundamental strings wrapping around the base ${\IP}^1$ which give the degree $k$ worldsheet instanton for this base whose K\"ahler modulus gives the dynamical scale $\Lambda$.\vspace{0.5em}

\noindent\underline{\bf Type IIA theory $\to$ M-theory:}\ \
Lifting to the M-theory leads to the relation (\ref{KNek_Stringy_rel}) between the K-theoretic $SU(N)$ instanton partition function with five dimensional Chern-Simons term and the classical stringy $SU(N)$ instanton partition function in four dimensions. We see that by this five dimensional lift, the FI parameter $\zeta$ is quantized, and it gives a five dimensional Chern-Simons coefficient $m\in{\IZ}$. As the result, we argue that the K-theoretic classical stringy instanton partition function is given by the K-theoretic instanton partition function (\ref{K_Nek_CS}) with five dimensional Chern-Simons coefficient.

The K-theoretic instanton partition function and the refined topological A-model partition function do not capture the quantum stringy corrections. In the next section, we study these quantum corrections.

\section{Quantum stringy corrections}\label{sec:quant_st_inst}

In the general Omega background, the stringy instanton partition function (\ref{sinst}) has the quantum stringy ($\alpha'$) corrections, and it gives the quantum-corrected K\"ahler potential $K_{k,N}$ on the K\"ahler moduli space of the ADHM moduli space ${\cal M}_{k,N}$ \cite{Jockers:2012dk, Gomis:2012wy, Bonelli:2013rja, Bonelli:2013mma}:
\begin{equation}
e^{-K_{k,N}(\epsilon_1,\epsilon_2,\vec{a},z)}=Z_{k,N}(\epsilon_1,\epsilon_2,\vec{a},z)
\end{equation}
up to an ambiguity by K\"ahler transformations $K_{k,N}(z,{\overline z})\to K_{k,N}(z,{\overline z})+f(z)+\overline{f(z)}$ where $f(z)$ is a holomorphic function of $z$. In the following, let us fix this ambiguity by a normalization
\begin{equation}
Z_{k,N}^{\mbox{\scriptsize norm}}(\epsilon_1,\epsilon_2,\vec{a},z)=(z\overline{z})^{-ir\frac{k}{N}\sum_{\alpha=1}^Na_{\alpha}}\bigg(\frac{\Gamma(-ir\epsilon_1)\Gamma(-ir\epsilon_2)}{\Gamma(ir\epsilon_1)\Gamma(ir\epsilon_2)}\bigg)^{kN}Z_{k,N}(\epsilon_1,\epsilon_2,\vec{a},z).
\label{sinst_norm}
\end{equation}
Here the first normalization factor $(z\overline{z})^{-ir\frac{k}{N}\sum_{\alpha=1}^Na_{\alpha}}$ shifts the classical stringy corrections, and then we see that this normalized partition function is invariant under simultaneous constant shift $a_{\alpha}\to a_{\alpha}+c$, $\alpha=1,\ldots,N$. As the result, the $U(1)$ factor is decoupled from the partition function: $U(N) \to SU(N)$.
By (\ref{Gamma_zeta}), the second normalization factor removes the dependence of the Euler constant $\gamma$ \cite{Bonelli:2013rja}, and fixes the precoefficients of the simple zeta values $\zeta(s)$ which receive the perturbative $\alpha'$ corrections. These normalizations fix the regularization scheme in the perturbative $\alpha'$ corrections mentioned in Section \ref{subsec:st_corr}. In this section, we study the structure of this partition function for $N=1,2,3$ cases, and find a universal structure of the quantum stringy corrections for arbitrary instanton charge $k$ as in (\ref{N1stringy_count}), (\ref{N2stringy_count}), and (\ref{N3stringy_count}).

\subsection{Stringy $U(1)$ instanton counting}\label{subsec:stringy_u1}

In the $N=1$ case, the ADHM moduli space ${\cal M}_{k,1}$ is isomorphic to the Hilbert scheme of points $\mbox{Hilb}_k({\IC}^2)$ on ${\IC^2}$. In this case, an equivariant mirror map described in (\ref{equiv_mirror1}) is needed for reproducing the correct Gromov-Witten invariants of $\mbox{Hilb}_k({\IC}^2)$, and thus we study \cite{Bonelli:2013rja}
\begin{equation}
\widehat{Z}_{k,1}^{\scriptsize \mbox{norm}}(\epsilon_1,\epsilon_2,z)=
(1+z)^{ikr\epsilon}(1+\overline{z})^{ikr\epsilon}Z_{k,1}^{\scriptsize \mbox{norm}}(\epsilon_1,\epsilon_2,-z).
\end{equation}
Note that the Coulomb modulus $a$ has been removed by the first normalization factor in (\ref{sinst_norm}).
We find that its asymptotic expansion around $r=0$ is given by
\begin{align}
\widehat{Z}_{k,1}^{\scriptsize \mbox{norm}}(\epsilon_1,\epsilon_2,z)&=
\sum_{\ell=0}^{\infty}\frac{(-1)^{\ell}G_{k,\ell}}{(\varepsilon_1\varepsilon_2)^{k-\ell}}l^{2\ell}(z)\nonumber\\
&
-\frac{G_{k-2,0}\varepsilon}{(\varepsilon_1\varepsilon_2)^{k-1}}\Big[e^{-K_1^{(3)}(z)}
+\varepsilon e^{-K_1^{(4)}(z)}
+\varepsilon^2e^{-K_1^{(5)}(z)}
+\varepsilon^3e^{-K_1^{(6)}(z)}\Big]\nonumber\\
&
+\frac{9G_{k-3,0}\varepsilon}{(\varepsilon_1\varepsilon_2)^{k-2}}\Big[e^{-K_{1;1}^{(5)}(z)}+\varepsilon e^{-K_{1;1}^{(6)}(z)}\Big]
+\frac{2G_{k-4,0}\varepsilon}{(\varepsilon_1\varepsilon_2)^{k-2}}\Big[e^{-K_{1;2}^{(5)}(z)}+\varepsilon e^{-K_{1;2}^{(6)}(z)}\Big]\nonumber\\
&
+\frac{2G_{k,2}\varepsilon}{(\varepsilon_1\varepsilon_2)^{k-2}}\Big[e^{-K_{1;3}^{(5)}(z)}+\varepsilon e^{-K_{1;3}^{(6)}(z)}\Big]+\epsilon\times{\cal O}(r^{-2k+7}),
\label{N1stringy_count}
\end{align}
where $\varepsilon_{1,2}=ir\epsilon_{1,2}$, $\varepsilon=\varepsilon_1+\varepsilon_2=ir\epsilon$, and
\begin{equation}
l^n(z)\equiv \frac{1}{n!}\log^{n}z\overline{z}.
\end{equation}
The coefficients $e^{-K_1^{(n)}(z)}, n=3,\ldots,6$, and $e^{-K_{1;i}^{(n)}(z)}, n=5,6, i=1,2,3$ are given below. We have checked this expansion up to 
$k=5$(-instanton), and for the classical stringy corrections up to $k=7$(-instanton). Here
\begin{equation}
G_{k\ge 0,\ell}=\sum_{|\mu|=k}\frac{1}{\prod_{s\in \mu}h_{\mu}(s)^2}\Big(\sum_{s\in \mu}c_{\mu}(s)\Big)^{2\ell},\ \ \ \ G_{k<0,\ell}\equiv 0,
\label{N1_Gk_classic}
\end{equation}
where $c_{\mu}(s)=\sum_{s\in \mu}(a_{\mu}(s)-\ell_{\mu}(s))=\sum_{(i,j)\in\mu}(j-i)$, and $h_{\mu}(s)=a_{\mu}(s)+\ell_{\mu}(s)+1$ is the hook length. Some of these numbers are given by
\begin{align}
&
G_{k,0}=\frac{1}{k!},\ \ \ \
G_{k,1}=\frac{k(k-1)}{2k!},\ \ \ \
G_{k,2}=\frac{k(k-1)(3k^2+k-12)}{2^2k!},\nonumber\\
&
G_{k,3}=\frac{k(k-1)(15k^4+30k^3-105k^2-700k+1344)}{2^3k!}.
\label{N1Gexplicit}
\end{align}
One finds that $G_{k,\ell}$ coincides with the equivariant classical intersection number $\langle D^{2\ell}\rangle^{\mbox{\scriptsize cl}}\big|_{\epsilon=0}$ of the divisor class $D$ in the $T$-equivariant cohomology $H_T^*\big(\mbox{Hilb}_k({\IC}^2),{\IQ}\big)$ computed in (\ref{class_intN1}), where $T=U(1)^2$ acts on ${\IC}^2$. As discussed in Appendix \ref{app:st_inst_hurwitz}, these numbers also give the disconnected simple Hurwitz numbers of ${\IP}^1$ which count degree $k$ covers of ${\IP}^1$ only with simple branch points (see (\ref{U1_inst_disHur})).

The normalized stringy $U(1)$ instanton partition function $\widehat{Z}_{k,1}^{\scriptsize \mbox{norm}}$ gives the K\"ahler potential on the quantum K\"ahler moduli space of $\mbox{Hilb}_k({\IC}^2)$, and by (\ref{Iz_ZV}) the non-perturbative $\alpha'$ corrections should be given by the equivariant Gromov-Witten invariants. In the following, we explicitly describe the expansion coefficients of (\ref{N1stringy_count}), and confirm that such corrections are given by the equivariant Gromov-Witten invariants computed in Appendix \ref{app:equiv_hilb}.

The coefficients $e^{-K_1^{(n)}(z)}$, $n=3,\ldots,6$ are given by
\begin{align}
e^{-K_1^{(3)}(z)}&=
\frac{1}{2}l^3(z)-3\zeta_3+\big(L_2(z)+c.c.\big)l^1(z)-2\big(L_3(z)+c.c.\big),\\
e^{-K_1^{(4)}(z)}&=
\frac{1}{2}l^4(z)-2\zeta_3l^1(z)+\big(L_2(z)+c.c.\big)l^2(z)
-\big(\widetilde{L}_3(z)+c.c.\big)l^1(z)-\big(L_2(z)-c.c.\big)^2,\\
e^{-K_1^{(5)}(z)}&=
\frac{1}{2}l^5(z)-2\zeta_3l^2(z)-4\zeta_3\big(L_2(z)+c.c.\big)-7\zeta_5+\big(L_2(z)+c.c.\big)l^3(z)\nonumber\\
&
-\big(\widetilde{L}_3(z)+c.c.\big)l^2(z)+\Big[\Big(\int\frac{dz}{z}\widetilde{L}_3(z)+\frac{2}{3}\int\frac{dz}{z}L_1(z)^3+c.c.\Big)
\nonumber\\
&
+2L_2(z)L_2(\overline{z})\Big]l^1(z)-2\Big[\Big(\int\int\frac{dz^2}{z^2}\widetilde{L}_3(z)+\frac{2}{3}\int\int\frac{dz^2}{z^2}L_1(z)^3-\int\frac{dz}{z}L_2(z)^2\nonumber\\
&
+2\int\frac{dz}{z}L_1(z)^2L_2(z)+c.c.\Big)+\big(L_2(z)+c.c.\big)\big(\widetilde{L}_3(z)+c.c.\big)\Big],
\end{align}
and
\begin{align}
e^{-K_1^{(6)}(z)}&=
\frac{1}{2}l^6(z)-2\zeta_3l^3(z)-4\zeta_3\big(L_2(z)+c.c.\big)l^1(z)+4\zeta_3\big(\widetilde{L}_3(z)+c.c.\big)+4\zeta_3^2-6\zeta_5l^1(z)\nonumber\\
&
+\big(L_2(z)+c.c.\big)l^4(z)-\big(\widetilde{L}_3(z)+c.c.\big)l^3(z)+\Big[\Big(\int\frac{dz}{z}\widetilde{L}_3(z)+\frac{2}{3}\int\frac{dz}{z}L_1(z)^3+c.c.\Big)
\nonumber\\
&
+2L_2(z)L_2(\overline{z})\Big]l^2(z)-\Big[\int\int\frac{dz^2}{z^2}\widetilde{L}_3(z)+\frac{2}{3}\int\int\frac{dz^2}{z^2}L_1(z)^3-\frac13\int\frac{dz}{z}L_1(z)^4\nonumber\\
&
+2L_2(z)\widetilde{L}_3(\overline{z})+c.c.\Big]l^1(z)+\big(\widetilde{L}_3(z)+c.c.\big)^2-2\big(L_2(z)-c.c.\big)\Big(\int\frac{dz}{z}\widetilde{L}_3(z)\nonumber\\
&
+\frac{2}{3}\int\frac{dz}{z}L_1(z)^3-c.c.\Big),
\end{align}
where
\begin{equation}
\widetilde{L}_3(z)\equiv L_3(z)-\int\frac{dz}{z}L_1(z)^2.
\end{equation}
Note that in the perturbative $\alpha'$ corrections which depend on the regularization scheme, the simple zeta values $\zeta_s\equiv\zeta(s)$ are involved. We see that the non-perturbative $\alpha'$ corrections are completely given in terms of the polylogarithms
\begin{equation}
L_0(z)\equiv{\rm Li}_0(z)=\frac{z}{1-z},\ \ \ \ 
L_{n+1}(z)\equiv{\rm Li}_{n+1}(z)=\int\frac{dz}{z}{\rm Li}_n(z).
\end{equation}
Especially $K_1^{(3)}(z)$ and $K_1^{(4)}(z)$ have the forms of the K\"ahler potentials on the K\"ahler moduli spaces of Calabi-Yau threefolds (\ref{CY3_Kahler}) and of Calabi-Yau fourfolds (\ref{CY4_Kahler}), respectively. Combining with the precoefficients in (\ref{N1stringy_count}) we argue that the non-perturbative $\alpha'$ corrections in $K_1^{(n)}(z)$ are determined by the equivariant three-point function on $\mbox{Hilb}_k({\IC}^2)$ computed in (\ref{DDD}),
\begin{equation}
K_1^{(n)}(z): \ \langle \epsilon^{n-3}D,D,D\rangle
=-\frac{k(k-1)\epsilon^{n-2}}{k!(\epsilon_1\epsilon_2)^{k-1}}\Big(\frac12+L_0(z)\Big),\ \ \ \ n=3,\ldots,6,
\label{N1DDD}
\end{equation}
where $D$ is the observable corresponding to the Poincar\'e dual of the divisor class in $H_T^*\big(\mbox{Hilb}_k({\IC}^2),{\IQ}\big)$.

The coefficients $e^{-K_{1;1}^{(5)}(z)}$ and $e^{-K_{1;2}^{(5)}(z)}$ are given by
\begin{align}
e^{-K_{1;1}^{(5)}(z)}&=
l^5(z)+\frac{16}{9}\zeta_3l^2(z)+\frac{320}{27}\zeta_5+\big(M_3(z)+c.c.\big)l^2(z)-3\big(M_4(z)+c.c.\big)l^1(z)\nonumber\\
&
+6\big(M_5(z)+c.c.\big),\\
e^{-K_{1;2}^{(5)}(z)}&=
\frac{1}{2}l^5(z)+\frac{3}{2}\zeta_3l^2(z)+\frac{45}{4}\zeta_5+\big(L_3(z)+c.c.\big)l^2(z)-3\big(L_4(z)+c.c.\big)l^1(z)\nonumber\\
&
+6\big(L_5(z)+c.c.\big).
\end{align}
The non-perturbative $\alpha'$ corrections in these coefficients are given in terms of
\begin{equation}
M_0(z)\equiv\frac{z(1-2z)}{1-z+z^2},\ \ \ \
M_{n+1}(z)\equiv\int\frac{dz}{z}M_n(z)
\end{equation} 
for $K_{1;1}^{(5)}(z)$, and $L_n(z)$ for $K_{1;2}^{(5)}(z)$. With the precoefficients in (\ref{N1stringy_count}) we argue that these quantum corrections are determined by the equivariant three-point functions computed in (\ref{DY1Y1}) for $K_{1;1}^{(5)}(z)$, and (\ref{DY2Y2}) for $K_{1;2}^{(5)}(z)$,
\begin{align}
&
K_{1;1}^{(5)}(z):\ \langle D,Y_2^{(1)},Y_2^{(1)}\rangle
=\frac{k(k-1)(k-2)\epsilon}{k!(\epsilon_1\epsilon_2)^{k-2}}\Big(1+M_0(z)\Big),\\
&
K_{1;2}^{(5)}(z):\ \langle D,Y_2^{(2)},Y_2^{(2)}\rangle
=\frac{k(k-1)(k-2)(k-3)\epsilon}{2k!(\epsilon_1\epsilon_2)^{k-2}}\Big(\frac12+L_0(z)\Big).
\label{N1DY2Y2}
\end{align}
Here $Y_2^{(1)}$ and $Y_2^{(2)}$ are the two independent observables corresponding to the Poincar\'e dual of (complex) codimension two cycle classes in $H_T^*\big(\mbox{Hilb}_k({\IC}^2),{\IQ}\big)$.

The coefficients $e^{-K_{1;1}^{(6)}(z)}$ and $e^{-K_{1;2}^{(6)}(z)}$ are given by
\begin{align}
e^{-K_{1;1}^{(6)}(z)}&
=5l^6(z)+\frac{10}{9}\zeta_3l^3(z)+\frac{32}{9}\zeta_3\big(L_2(z)+c.c.\big)l^1(z)-\frac{2}{9}\zeta_3\big(32L_3(z)+3M_3(z)+c.c.\big)\nonumber\\
&
-\frac{160}{27}\zeta_3^2+\frac{106}{9}\zeta_5l^1(z)+2\big(L_2(z)+c.c.\big)l^4(z)-4\big(L_3(z)-M_3(z)+c.c.\big)l^3(z)\nonumber\\
&
+2\Big[3L_4(z)-4M_4(z)-\int\int\frac{dz^2}{z^2}L_1(z)M_1(z)+\frac34\int\int\frac{dz^2}{z^2}M_1(z)^2\nonumber\\
&
+\int\frac{dz}{z}L_1(z)M_2(z)+c.c.\Big]l^2(z)-2\Big[\Big(3L_5(z)-4M_5(z)\nonumber-\int\int\int\frac{dz^3}{z^3}L_1(z)M_1(z)\\
&
+\frac34\int\int\int\frac{dz^3}{z^3}M_1(z)^2+\int\int\frac{dz^2}{z^2}L_1(z)M_2(z)+3\int\frac{dz}{z}L_1(z)M_3(z)\nonumber\\
&
+\frac34\int\frac{dz}{z}M_2(z)^2+c.c.\Big)-\big(L_2(z)+c.c.\big)\big(M_3(z)+c.c.\big)\Big]l^1(z)-\frac{1}{2}\big(M_3(z)\nonumber\\
&
+c.c.\big)\big(8L_3(z)-3M_3(z)+c.c.\big)+6\big(L_2(z)-c.c.\big)\big(M_4(z)-c.c.\big),
\end{align}
and
\begin{align}
e^{-K_{1;2}^{(6)}(z)}&
=2l^6(z)+\frac{3}{2}\zeta_3l^3(z)+3\zeta_3\big(L_2(z)+c.c.\big)l^1(z)-6\zeta_3\big(L_3(z)+c.c.\big)-\frac{9}{2}\zeta_3^2+12\zeta_5l^1(z)\nonumber\\
&
+\big(L_2(z)+c.c.\big)l^4(z)+\big(L_3(z)+c.c.\big)l^3(z)-\big(3L_4(z)-L_2(z)^2+c.c.\big)l^2(z)\nonumber\\
&
+\Big[3L_5(z)+3\int\frac{dz}{z}L_2(z)^2-2L_2(z)\big(2L_3(z)-L_3(\overline{z})\big)+c.c.\Big]l^1(z)\nonumber\\
&
-2\big(L_3(z)+c.c.\big)^2+6\big(L_2(z)-c.c.\big)\big(L_4(z)-c.c.\big).
\end{align}
We argue that the non-perturbative $\alpha'$ corrections in these coefficients are determined by the equivariant three-point functions as
\begin{align}
&
K_{1;1}^{(6)}(z):\ \langle \epsilon_1\epsilon_2\epsilon D,D,D\rangle\ \ \mbox{and}\ \ \langle \epsilon D,Y_2^{(1)},Y_2^{(1)}\rangle,\\
&
K_{1;2}^{(6)}(z):\ \langle \epsilon_1\epsilon_2\epsilon D,D,D\rangle\ \ \mbox{and}\ \ \langle \epsilon D,Y_2^{(2)},Y_2^{(2)}\rangle.
\end{align}

We find that the coefficients $K_{1;3}^{(5)}(z)$ and $K_{1;3}^{(6)}(z)$ are written in terms of the other coefficients as
\begin{align}
e^{-K_{1;3}^{(5)}(z)}&=
\frac13\Big(e^{-K_1^{(3)}(z)}l^2(z)-4e^{-K_{1;2}^{(5)}(z)}\Big)\sim l^5(z),\\
e^{-K_{1;3}^{(6)}(z)}&=
\frac13\Big(e^{-2K_1^{(3)}(z)}+e^{-K_1^{(4)}(z)}l^2(z)-4e^{-K_{1;2}^{(6)}(z)}\Big)\sim \frac32l^6(z).
\end{align}
The stringy $k=2$-instanton partition function in (\ref{N1stringy_count}) receives these non-perturbative $\alpha'$ corrections, whereas the equivariant three-point functions $\langle D,Y_2^{(1)},Y_2^{(1)}\rangle$, $\langle D,Y_2^{(2)},Y_2^{(2)}\rangle$ computed in (\ref{DY1Y1}), (\ref{DY2Y2}) are trivial for $k=2$, and thus we argue that the non-perturbative $\alpha'$ corrections in each coefficient are determined by the equivariant three-point functions as
\begin{equation}
K_{1;3}^{(5)}(z):\ \langle \epsilon_1\epsilon_2 D,D,D\rangle,\ \ \ \
K_{1;3}^{(6)}(z):\ \langle \epsilon_1\epsilon_2\epsilon D,D,D\rangle.
\end{equation}

We also find that the classical intersection numbers obtained by the above computation completely coincide with the equivariant classical intersection numbers (\ref{class_intN1}) of the divisor class $D$ in $H_T^*\big(\mbox{Hilb}_k({\IC}^2),{\IQ}\big)$.

\subsection{Stringy $U(2)$ instanton counting}\label{subsec:stringy_u2}

In the $N=2$ case, let us consider a normalized partition function on ${\cal M}_{k,2}$ \cite{Bonelli:2013rja}:
\begin{equation}
\widehat{Z}_{k,2}^{\scriptsize \mbox{norm}}(\epsilon_1,\epsilon_2,\vec{a},z)=
(z\overline{z})^{i\frac{k}{2}r\epsilon}Z_{k,2}^{\scriptsize \mbox{norm}}(\epsilon_1,\epsilon_2,\vec{a},z).
\label{N2stringy_norm}
\end{equation}
This normalization shifts $a_{\alpha}$ to $a_{\alpha}+\epsilon/2$.
In the following, by putting $\zeta(s)=0$ we ignore the perturbative $\alpha'$ corrections which depend on the second normalization factor in (\ref{sinst_norm}). Then in the asymptotic expansion around $r=0$, we only need to consider the leading term of $Z_{\mbox{\scriptsize L}}$ in (\ref{sinst_loop}). Up to $k=4$(-instanton), we find that the asymptotic expansion has the form
\begin{align}
\widehat{Z}_{k,2}^{\scriptsize \mbox{norm}}(\epsilon_1,\epsilon_2,\vec{a},z)&=
\sum_{\ell=0}^{3}(-1)^{\ell}G_{k,\ell}^{(2)}(\varepsilon_1,\varepsilon_2,\vec{\mathfrak a})l^{2\ell}(z)\nonumber\\
&
-\frac{\varepsilon}{\varepsilon_1\varepsilon_2}G_{k-1,0}^{(2)}(\varepsilon_1,\varepsilon_2,\vec{\mathfrak a})\Big[e^{-K_2^{(3)}(z)}
+\varepsilon e^{-K_2^{(4)}(z)}
+\varepsilon^2e^{-K_2^{(5)}(z)}+\varepsilon^3e^{-K_2^{(6)}(z)}\Big]\nonumber\\
&
+\frac{2\varepsilon}{\varepsilon_1^2\varepsilon_2^2}G_{k-2,0}^{(2)}(\varepsilon_1,\varepsilon_2,\vec{\mathfrak a})\Big[e^{-K_{2;2}^{(5)}(z)}+\varepsilon e^{-K_{2;2}^{(6)}(z)}\Big]\nonumber\\
&
+2\varepsilon G_{k,2}^{(2)}(\varepsilon_1,\varepsilon_2,\vec{\mathfrak a})\Big[e^{-K_{2;3}^{(5)}(z)}+\varepsilon e^{-K_{2;3}^{(6)}(z)}\Big]+{\cal O}(r^{-4k+7}),
\label{N2stringy_count}
\end{align}
where $\varepsilon_{1,2}=ir\epsilon_{1,2}$, $\varepsilon=\varepsilon_1+\varepsilon_2$, $\vec{\mathfrak a}=ir\vec{a}$, and $l^n(z)=\frac{1}{n!}\log^{n}z\overline{z}$. We see that by the normalization factor $(z\overline{z})^{i\frac{k}{2}r\epsilon}$ in (\ref{N2stringy_norm}), the odd power terms with $l^{2\ell+1}$ for $\epsilon=0$ have been removed from the first line of (\ref{N2stringy_count}).
Here $G_{k,\ell}^{(2)}$ behaves as ${\cal O}(r^{-4k+2\ell})$, and its leading term is given by the Nekrasov partition function (\ref{Nekrasov_in}):
\begin{equation}
G_{k,0}^{(2)}(\varepsilon_1,\varepsilon_2,\vec{\mathfrak a})=Z_{k,2}^{\mbox{\scriptsize Nek}}(\varepsilon_1,\varepsilon_2,\vec{\mathfrak a}),\ \ \ \
Z_{k<0,2}^{\mbox{\scriptsize Nek}}(\varepsilon_1,\varepsilon_2,\vec{\mathfrak a})\equiv 0.
\end{equation}
Up to $k=4$, one has $Z_{0,2}^{\mbox{\scriptsize Nek}}(\varepsilon_1,\varepsilon_2,\vec{\mathfrak a})=1$,
\begin{align}
&
Z_{1,2}^{\mbox{\scriptsize Nek}}(\varepsilon_1,\varepsilon_2,\vec{\mathfrak a})=\frac{2}{\varepsilon_1\varepsilon_2D_{1,1}},\\
&
Z_{2,2}^{\mbox{\scriptsize Nek}}(\varepsilon_1,\varepsilon_2,\vec{\mathfrak a})=\frac{8\varepsilon_1^2+17\varepsilon_1\varepsilon_2+8\varepsilon_2^2-2\mathfrak{a}^2}{\varepsilon_1^2\varepsilon_2^2D_{1,1}D_{1,2}D_{2,1}},\\
&
Z_{3,2}^{\mbox{\scriptsize Nek}}(\varepsilon_1,\varepsilon_2,\vec{\mathfrak a})=\frac{N_3}{3\varepsilon_1^3\varepsilon_2^3D_{1,1}D_{1,2}D_{2,1}D_{1,3}D_{3,1}},\\
&
Z_{4,2}^{\mbox{\scriptsize Nek}}(\varepsilon_1,\varepsilon_2,\vec{\mathfrak a})=\frac{N_4}{6\varepsilon_1^4\varepsilon_2^4D_{1,1}D_{1,2}D_{2,1}D_{1,3}D_{3,1}D_{2,2}D_{1,4}D_{4,1}}.
\end{align}
Here $\mathfrak{a}=\mathfrak{a}_1-\mathfrak{a}_2$, and $D_{i,j}=D_{i,j}(\varepsilon_1,\varepsilon_2,\vec{\mathfrak a})$, $N_{3}=N_{3}(\varepsilon_1,\varepsilon_2,\vec{\mathfrak a})$, $N_{4}=N_{4}(\varepsilon_1,\varepsilon_2,\vec{\mathfrak a})$ are given by
\begin{align}
&
D_{i,j}=(i\varepsilon_1+j\varepsilon_2)^2-\mathfrak{a}^2,\\
&
N_3=2\big(72(\varepsilon_1^4+\varepsilon_2^4)+363(\varepsilon_1^3\varepsilon_2+\varepsilon_1\varepsilon_2^3)+594\varepsilon_1^2\varepsilon_2^2-(26\varepsilon_1^2+47\varepsilon_1\varepsilon_2+26\varepsilon_2^2)\mathfrak{a}^2+2\mathfrak{a}^4\big),\\
&
N_4=
9216(\varepsilon_1^8+\varepsilon_2^8)+100608(\varepsilon_1^7\varepsilon_2+\varepsilon_1\varepsilon_2^7)+440688(\varepsilon_1^6\varepsilon_2^2+\varepsilon_1^2\varepsilon_2^6)+1009131(\varepsilon_1^5\varepsilon_2^3+\varepsilon_1^3\varepsilon_2^5)\nonumber\\
&\hspace{2em}
+1319994\varepsilon_1^4\varepsilon_2^4-\big(6208(\varepsilon_1^6+\varepsilon_2^6)+44336(\varepsilon_1^5\varepsilon_2+\varepsilon_1\varepsilon_2^5)+124139(\varepsilon_1^4\varepsilon_2^2+\varepsilon_1^2\varepsilon_2^4)\nonumber\\
&\hspace{2em}
+171845\varepsilon_1^3\varepsilon_2^3\big)\mathfrak{a}^2+\big(1440(\varepsilon_1^4+\varepsilon_2^4)+5644(\varepsilon_1^3\varepsilon_2+\varepsilon_1\varepsilon_2^3)+8651\varepsilon_1^2\varepsilon_2^2\big)\mathfrak{a}^4\nonumber\\
&\hspace{2em}
-\big(132(\varepsilon_1^2+\varepsilon_2^2)+212\varepsilon_1\varepsilon_2\big)\mathfrak{a}^6+4\mathfrak{a}^8.
\end{align}
We find that $G_{k,1}^{(2)}$ is also given by the Nekrasov partition function
\begin{equation}
G_{k,1}^{(2)}(\varepsilon_1,\varepsilon_2,\vec{\mathfrak a})=\frac{1}{2\varepsilon_1\varepsilon_2}Z_{k-1,2}^{\mbox{\scriptsize Nek}}(\varepsilon_1,\varepsilon_2,\vec{\mathfrak a}).
\end{equation}
Up to $k=4$, $G_{k,2}^{(2)}$ are given by
\begin{align}
G_{1,2}^{(2)}(\varepsilon_1,\varepsilon_2,\vec{\mathfrak a})&=
\frac{\varepsilon^2-\mathfrak{a}^2}{8\varepsilon_1\varepsilon_2},\\
G_{2,2}^{(2)}(\varepsilon_1,\varepsilon_2,\vec{\mathfrak a})&=
\frac{\varepsilon_1^2+3\varepsilon_1\varepsilon_2+\varepsilon_2^2-\mathfrak{a}^2}{\varepsilon_1^2\varepsilon_2^2D_{1,1}},\\
G_{3,2}^{(2)}(\varepsilon_1,\varepsilon_2,\vec{\mathfrak a})&=\frac{N_{3,2}}{8\varepsilon_1^3\varepsilon_2^3D_{1,1}D_{1,2}D_{2,1}},\\
G_{4,2}^{(2)}(\varepsilon_1,\varepsilon_2,\vec{\mathfrak a})&=\frac{N_{4,2}}{3\varepsilon_1^4\varepsilon_2^4D_{1,1}D_{1,2}D_{2,1}D_{1,3}D_{3,1}},
\end{align}
where
\begin{align}
N_{3,2}&=56(\varepsilon_1^4+\varepsilon_2^4)+337(\varepsilon_1^3\varepsilon_2+\varepsilon_1\varepsilon_2^3)+582\varepsilon_1^2\varepsilon_2^2-(70\varepsilon_1^2+133\varepsilon_1\varepsilon_2+70\varepsilon_2^2)\mathfrak{a}^2+14\mathfrak{a}^4,\\
N_{4,2}&=
180(\varepsilon_1^6+\varepsilon_2^6)+1767(\varepsilon_1^5\varepsilon_2+\varepsilon_1\varepsilon_2^5)
+6018(\varepsilon_1^4\varepsilon_2^2+\varepsilon_1^2\varepsilon_2^4)+8936\varepsilon_1^3\varepsilon_2^3-\big(245(\varepsilon_1^4+\varepsilon_2^4)\nonumber\\
&
+1062(\varepsilon_1^3\varepsilon_2+\varepsilon_1\varepsilon_2^3)
+1676\varepsilon_1^2\varepsilon_2^2\big)\mathfrak{a}^2+\big(70(\varepsilon_1^2+\varepsilon_2^2)+111\varepsilon_1\varepsilon_2\big)\mathfrak{a}^4-5\mathfrak{a}^6.
\end{align}
Up to $k=4$, $G_{k,3}^{(2)}$ are given by
\begin{align}
G_{1,3}^{(2)}(\varepsilon_1,\varepsilon_2,\vec{\mathfrak a})&=
\frac{(\varepsilon^2-\mathfrak{a}^2)^2}{32\varepsilon_1\varepsilon_2},\\
G_{2,3}^{(2)}(\varepsilon_1,\varepsilon_2,\vec{\mathfrak a})&=
\frac{N_{2,3}}{\varepsilon_1^2\varepsilon_2^2D_{1,1}},\\
G_{3,3}^{(2)}(\varepsilon_1,\varepsilon_2,\vec{\mathfrak a})&=\frac{N_{3,3}}{32\varepsilon_1^3\varepsilon_2^3D_{1,1}D_{1,2}D_{2,1}},\\
G_{4,3}^{(2)}(\varepsilon_1,\varepsilon_2,\vec{\mathfrak a})&=\frac{N_{4,3}}{3\varepsilon_1^4\varepsilon_2^4D_{1,1}D_{1,2}D_{2,1}D_{1,3}D_{3,1}},
\end{align}
where
\begin{align}
N_{2,3}&=(\varepsilon_1^4+\varepsilon_2^4)+10(\varepsilon_1^3\varepsilon_2+\varepsilon_1\varepsilon_2^3)+19\varepsilon_1^2\varepsilon_2^2-2(\varepsilon_1^2+5\varepsilon_1\varepsilon_2+\varepsilon_2^2)\mathfrak{a}^2+\mathfrak{a}^4,\\
N_{3,3}&=
488(\varepsilon_1^6+\varepsilon_2^6)+5905(\varepsilon_1^5\varepsilon_2+\varepsilon_1\varepsilon_2^5)
+22244(\varepsilon_1^4\varepsilon_2^2+\varepsilon_1^2\varepsilon_2^4)+34310\varepsilon_1^3\varepsilon_2^3-\big(1098(\varepsilon_1^4+\varepsilon_2^4)\nonumber\\
&
+7754(\varepsilon_1^3\varepsilon_2+\varepsilon_1\varepsilon_2^3)
+12392\varepsilon_1^2\varepsilon_2^2\big)\mathfrak{a}^2+\big(732(\varepsilon_1^2+\varepsilon_2^2)+1849\varepsilon_1\varepsilon_2\big)\mathfrak{a}^4-122\mathfrak{a}^6,\\
N_{4,3}&=
612(\varepsilon_1^8+\varepsilon_2^8)+9867(\varepsilon_1^7\varepsilon_2+\varepsilon_1\varepsilon_2^7)+57891(\varepsilon_1^6\varepsilon_2^2+\varepsilon_1^2\varepsilon_2^6)+160209(\varepsilon_1^5\varepsilon_2^3+\varepsilon_1^3\varepsilon_2^5)\nonumber\\
&
+224058\varepsilon_1^4\varepsilon_2^4-\big(1445(\varepsilon_1^6+\varepsilon_2^6)+14434(\varepsilon_1^5\varepsilon_2+\varepsilon_1\varepsilon_2^5)+45445(\varepsilon_1^4\varepsilon_2^2+\varepsilon_1^2\varepsilon_2^4)\nonumber\\
&
+65416\varepsilon_1^3\varepsilon_2^3\big)\mathfrak{a}^2+\big(1071(\varepsilon_1^4+\varepsilon_2^4)+5003(\varepsilon_1^3\varepsilon_2+\varepsilon_1\varepsilon_2^3)+7729\varepsilon_1^2\varepsilon_2^2\big)\mathfrak{a}^4\nonumber\\
&
-\big(255(\varepsilon_1^2+\varepsilon_2^2)+436\varepsilon_1\varepsilon_2\big)\mathfrak{a}^6+17\mathfrak{a}^8.
\end{align}
Up to the precoefficients of $\zeta(s)$ given by the ratio of two homogeneous polynomials of $\varepsilon_{1,2}$ and $\mathfrak{a}$ which depend on the regularization scheme, we find that the coefficients in (\ref{N2stringy_count}) coincide with the coefficients in (\ref{N1stringy_count}) for the stringy $U(1)$ instanton partition function:
\begin{equation}
e^{-K_2^{(n)}(z)}\equiv e^{-K_1^{(n)}(z)},\ \ \ \ 
e^{-K_{2;2}^{(d)}(z)}\equiv e^{-K_{1;2}^{(d)}(z)},\ \ \ \ 
e^{-K_{2;3}^{(d)}(z)}\equiv e^{-K_{1;3}^{(d)}(z)},
\end{equation}
where $n=3,\ldots,6$ and $d=5,6$.

By combining the above results with the results in Section \ref{subsec:stringy_u1}, we can conjecture the quantum corrections of equivariant three-point functions on ${\cal M}_{k,2}$ up to the normalization constants determined by the classical intersection numbers. Let $D$ be the Poincar\'e dual observable of the divisor class in the $T$-equivariant cohomology $H_T^*\big({\cal M}_{k,2}\big)$ of ${\cal M}_{k,2}$, where $T$ is the $2+2$ dimensional torus as in footnote 6 of Section \ref{subsec:st_corr}. From the precoefficient of $e^{-K_2^{(3)}(z)}$ in (\ref{N2stringy_count}) which does not depend on the normalization (\ref{N2stringy_norm}), by (\ref{N1DDD}) we find
\begin{equation}
\langle D,D,D\rangle
=-\frac{c_1\epsilon}{\epsilon_1\epsilon_2}Z_{k-1,2}^{\mbox{\scriptsize Nek}}(\varepsilon_1,\varepsilon_2,\vec{\mathfrak a})\Big(\frac12+L_0(z)\Big),
\label{DDD_U2_st}
\end{equation}
where $c_1$ is a constant, and $L_0(z)=z/(1-z)$. From the precoefficient of $e^{-K_{2;2}^{(5)}(z)}$ in (\ref{N2stringy_count}), by (\ref{N1DY2Y2}) we also find that there exists a cohomological observable $Y_2^{(2)}$ dual to a (complex) codimension two cycle class in $H_T^*\big({\cal M}_{k,2}\big)$ with the quantum corrections in three-point function as
\begin{equation}
\langle D,Y_2^{(2)},Y_2^{(2)}\rangle
=\frac{c_2\epsilon}{2\epsilon_1^2\epsilon_2^2}Z_{k-2,2}^{\mbox{\scriptsize Nek}}(\varepsilon_1,\varepsilon_2,\vec{\mathfrak a})\Big(\frac12+L_0(z)\Big),
\label{DYY_U2_st}
\end{equation}
where $c_2$ is a constant. Up to $k=4$, there are no other non-trivial Poincar\'e dual observables $Y$ of codimension two cycle classes with the quantum corrections in the three-point function $\langle D,Y,Y\rangle$.

In Appendix \ref{app:equiv_adhm}, we will describe the quantum cohomology ring $H_T^*\big({\cal M}_{k,N}\big)$ by the tensor product of Fock spaces \cite{Bar, Maulik:2012wi}, and by computing the quantum parts $\langle D, Y, Y\rangle^{\mbox{\scriptsize qu}}$ of three-point functions, we see that
the above results (\ref{DDD_U2_st}) and (\ref{DYY_U2_st}) are surely consistent with (\ref{DY1Y1_adhm}) and (\ref{DY2Y2_adhm}), respectively.

\subsection{Stringy $U(3)$ instanton counting}\label{subsec:stringy_u3}

In the $N=3$ case, as in the $N=2$ case we consider a normalization
\begin{equation}
\widehat{Z}_{k,3}^{\scriptsize \mbox{norm}}(\epsilon_1,\epsilon_2,\vec{a},z)=
(z\overline{z})^{i\frac{k}{2}r\epsilon}Z_{k,3}^{\scriptsize \mbox{norm}}(\epsilon_1,\epsilon_2,\vec{a},z),
\end{equation}
and ignore the perturbative $\alpha'$ corrections by putting $\zeta(s)=0$. As in Section \ref{subsec:stringy_u2}, this normalization shifts $a_{\alpha}$ to $a_{\alpha}+\epsilon/2$.
Up to $k=3$(-instanton), we find the asymptotic expansion around $r=0$ as
\begin{align}
\widehat{Z}_{k,3}^{\scriptsize \mbox{norm}}(\epsilon_1,\epsilon_2,\vec{a},z)&=
\sum_{\ell=0}^{3}(-1)^{\ell}G_{k,\ell}^{(3)}(\varepsilon_1,\varepsilon_2,\vec{\mathfrak a})l^{2\ell}(z)
-\sum_{\ell=0}^{2}(-1)^{\ell}H_{k,\ell}^{(3)}(\varepsilon_1,\varepsilon_2,\vec{\mathfrak a})l^{2\ell+1}(z)\nonumber\\
&
+\frac{\varepsilon}{\varepsilon_1\varepsilon_2}G_{k-1,0}^{(3)}(\varepsilon_1,\varepsilon_2,\vec{\mathfrak a})\Big[e^{-K_{3;2}^{(5)}(z)}+\varepsilon e^{-K_{3;2}^{(6)}(z)}\Big]\nonumber\\
&
-\frac{3\varepsilon}{\varepsilon_1\varepsilon_2}H_{k-1,0}^{(3)}(\varepsilon_1,\varepsilon_2,\vec{\mathfrak a})l^{1}(z)e^{-K_{3;2}^{(5)}(z)}+{\cal O}(r^{-6k+7}),
\label{N3stringy_count}
\end{align}
where $\varepsilon_{1,2}=ir\epsilon_{1,2}$, $\varepsilon=\varepsilon_1+\varepsilon_2$, $\vec{\mathfrak a}=ir\vec{a}$, and $l^n(z)=\frac{1}{n!}\log^{n}z\overline{z}$. Here $G_{k,\ell}^{(3)}$ behaves as ${\cal O}(r^{-6k+2\ell})$, and the leading term $G_{k,0}^{(3)}$ coincides with the Nekrasov partition function (\ref{Nekrasov_in}):
\begin{equation}
G_{k,0}^{(3)}(\varepsilon_1,\varepsilon_2,\vec{\mathfrak a})=Z_{k,3}^{\mbox{\scriptsize Nek}}(\varepsilon_1,\varepsilon_2,\vec{\mathfrak a}),\ \ \ \ 
Z_{k<0,3}^{\mbox{\scriptsize Nek}}(\varepsilon_1,\varepsilon_2,\vec{\mathfrak a})\equiv 0.
\end{equation}
For example, $Z_{0,3}^{\mbox{\scriptsize Nek}}(\varepsilon_1,\varepsilon_2,\vec{\mathfrak a})=1$ and
\begin{equation}
Z_{1,3}^{\mbox{\scriptsize Nek}}(\varepsilon_1,\varepsilon_2,\vec{\mathfrak a})=\frac{2(3\varepsilon^2+\mathfrak{a}_{12}\mathfrak{a}_{23}+\mathfrak{a}_{23}\mathfrak{a}_{31}+\mathfrak{a}_{31}\mathfrak{a}_{12})}{\varepsilon_1\varepsilon_2(\varepsilon^2-\mathfrak{a}_{12}^2)(\varepsilon^2-\mathfrak{a}_{23}^2)(\varepsilon^2-\mathfrak{a}_{31}^2)},
\end{equation}
where $\mathfrak{a}_{\alpha\beta}=\mathfrak{a}_{\alpha}-\mathfrak{a}_{\beta}$. $H_{k,\ell}^{(3)}$ ($H_{k\le 0,\ell}^{(3)}\equiv 0$) behaves as ${\cal O}(r^{-6k+2\ell+1})$, and for example
\begin{equation}
H_{1,0}^{(3)}=\frac{(\mathfrak{a}_{12}-\mathfrak{a}_{23})(\mathfrak{a}_{23}-\mathfrak{a}_{31})(\mathfrak{a}_{31}-\mathfrak{a}_{12})}{3\varepsilon_1\varepsilon_2(\varepsilon^2-\mathfrak{a}_{12}^2)(\varepsilon^2-\mathfrak{a}_{23}^2)(\varepsilon^2-\mathfrak{a}_{31}^2)}.
\end{equation}
Up to the perturbative $\alpha'$ corrections we find $e^{-K_{3;2}^{(5)}(z)}\equiv e^{-K_{1;2}^{(5)}(z)}\equiv e^{-K_{2;2}^{(5)}(z)}$, and the coefficient $e^{-K_{3;2}^{(6)}(z)}$ is given by
\begin{align}
e^{-K_{3;2}^{(6)}(z)}&\equiv
\frac{3}{4}l^6(z)+\frac32\big(L_3(z)+c.c.\big)l^3(z)-\frac32\Big(2L_4(z)-\int\int\frac{dz^2}{z^2}L_1(z)^2+c.c.\Big)l^2(z)\nonumber\\
&
+\frac32\Big(2L_5(z)-\int\int\int\frac{dz^3}{z^3}L_1(z)^2-\int\frac{dz}{z}L_2(z)^2+c.c.\Big)l^1(z)+\frac32\big(L_3(z)+c.c.\big)^2.
\end{align}

As in the $N=2$ case, we find that there exist the observable $D$ for the divisor class and an observable $Y_2^{(2)}$ for a (complex) codimension two dual cycle class in $H_T^*\big({\cal M}_{k,3}\big)$ with the quantum corrections in three-point function as
\begin{equation}
\langle D,Y_2^{(2)},Y_2^{(2)}\rangle
=\frac{c \epsilon}{\epsilon_1\epsilon_2}Z_{k-1,3}^{\mbox{\scriptsize Nek}}(\varepsilon_1,\varepsilon_2,\vec{\mathfrak a})\Big(\frac12+L_0(z)\Big),
\label{DYY_U3_st}
\end{equation}
where $c$ is a constant. Here $H_T^*\big({\cal M}_{k,3}\big)$ is the $T$-equivariant cohomology of ${\cal M}_{k,3}$ with the $3+2$ dimensional torus action $T$ mentioned in footnote 6. By the expansion (\ref{N3stringy_count}), we see that the three-point function $\langle D,D,D\rangle$ does not have quantum corrections.

As same as the $N=2$ case, we see that the above result (\ref{DYY_U3_st}) is consistent with the result (\ref{DY1Y1_adhm}) in Appendix \ref{app:equiv_adhm}.

\section{Conclusion and discussions}\label{sec:conclusion}

In this paper, we have studied the stringy instanton partition function of four dimensional ${\cal N}=2$ $U(N)$ supersymmetric gauge theory given in \cite{Bonelli:2013rja}. In Section \ref{subsec:rel_k_inst}, we found that the stringy instanton partition function whose $\alpha'$ corrections have been removed coincides with the four dimensional limit (\ref{4d_lim}) of the K-theoretic instanton partition function with a five dimensional Chern-Simons coefficient. Here the K\"ahler modulus $\zeta$ of the ADHM moduli space was obtained from the strong coupling limit of the five dimensional Chern-Simons term. We also discussed that the classical stringy instanton partition function is embedded to (refined) topological string theory on the local toric Calabi-Yau threefolds labeled by $m\in {\IZ}$ in Section \ref{subsec:rel_top_st}. This gives geometric engineering of the instantons with classical stringy corrections, and we provided a physical explanation of this realization in Section \ref{subsec:brane_geom}

We further studied the stringy instanton partition function with the $\alpha'$ corrections for $U(1)$, $U(2)$, and $U(3)$ cases. We found that the quantum stringy corrections have the universal structure for arbitrary instanton charge $k$, as in (\ref{N1stringy_count}), (\ref{N2stringy_count}), and (\ref{N3stringy_count}). This universal structure shows that the stringy corrections for an instanton charge $k$ contain the stringy corrections for instanton charges $l\le k$. From the viewpoint of the Fock space description of the equivariant cohomology ring of the ADHM moduli space ${\cal M}_{k,N}$ in Appendix \ref{app:equiv_hilb} and \ref{app:equiv_adhm}, such structure is clear
from the construction of the elements of the equivariant cohomology by the generators of the Heisenberg algebra. Note that this structure seems to be related with a compactification of the UV non-compactness by point-like instantons, which is called the Uhlenbeck compactification of the framed moduli space ${\cal M}^{(0)}_{k,N}$ of point-like instantons on $S^4\cong{\IC}^2 \cup \{\infty\}$ (see e.g. \cite{Nekrasov:2002qd, Nakajima:2003pg, Nakajima:2003uh}):\footnote{We would like to thank the referee for pointing out this similarity.}
$$
\widehat{\cal M}_{k,N}=\bigsqcup_{l=0}^k {\cal M}^{(0)}_{l,N} \times S^{k-l}{\IC}^2,
$$
where $S^l{\IC}^2$ is the $l$-th symmetric product of ${\IC}^2$. After resolving the orbifold singularities in $\widehat{\cal M}_{k,N}$, one obtains ${\cal M}_{k,N}$.

For $U(1)$ case, as discussed in \cite{Bonelli:2013rja} we read off some equivariant three-point functions on ${\cal M}_{k,1}\cong\mbox{Hilb}_k({\IC}^2)$, and confirmed the agreement with the computations in Appendix \ref{app:equiv_hilb}. Using this result, we extracted some equivariant three-point functions on ${\cal M}_{k,2}$ and ${\cal M}_{k,3}$ from the stringy instanton partition functions, and also checked that these results are consistent with the computations in Appendix \ref{app:equiv_adhm}.

It would be interesting to further study the classical and the quantum structure of the $U(N)$ stringy instanton partition function, and to compare the structure with the discussion of the quantum multiplication for the ADHM moduli space ${\cal M}_{k,N}$ (as described in Appendix \ref{app:equiv_adhm}) \cite{Maulik:2012wi}. The refined topological vertex \cite{Awata:2005fa, Iqbal:2007ii} do not capture the quantum stringy corrections, and thus it would be also interesting to formulate ``quantum refined topological vertex'' which captures such quantum corrections.

A six dimensional analogue of the four dimensional instanton partition function was discussed in \cite{Jafferis:2007sg, Cirafici:2008sn, Awata:2009dd}. It would be interesting to discuss the stringy generalization of the six dimensional instanton partition function.

\subsection*{Acknowledgements}

I would like to thank Rajesh Gopakumar, Yoshinori Honma, Kentaro Hori, and Piotr Su{\l}kowski for useful discussions and comments. I would also like to thank the organizers of National Strings Meeting 2013 held at IIT Kharagpur, where a part of this work was presented. Many of the results presented here were obtained during the period of postdoctoral fellow at Harish-Chandra Research Institute (HRI). I would like to thank HRI for supporting my work. I would also like to thank Kavli IPMU for hospitality during the final stage of this work. This work has been supported by the ERC Starting Grant no. 335739 ``Quantum fields and knot homologies'' funded by the European Research Council under the European Union's Seventh Framework Programme.

\appendix
\section{Multiple gamma function}\label{app:multi_gamma}

The multiple gamma function is defined by
\begin{equation}
\Gamma_r(x|\omega_1,\ldots,\omega_r)=\exp\left(\left.\frac{\partial}{\partial s}\right|_{s=0}\zeta_r(s,x|\omega_1,\ldots,\omega_r)\right),
\end{equation}
where $\zeta_r$ is the Barnes zeta function defined by the analytic continuation of an infinite sum
\begin{equation}
\zeta_r(s,x|\omega_1,\ldots,\omega_r)=\sum_{n_1,\ldots,n_r=0}^{\infty}\frac{1}{(x+n_1\omega_1+\cdots+n_r\omega_r)^s},\ \ \ \ 
{\rm Re}(s)>r.
\end{equation}

\subsection{Zeta function regularization}\label{subapp:zeta}

Using the multiple gamma function, one can regularize an infinite product by
\begin{equation}
\Big[\prod_{n_1,\ldots,n_r=0}^{\infty}(x+n_1\omega_1+\cdots+n_r\omega_r)^{-1}\Big]_{\mbox{\scriptsize reg}}=\Gamma_r(x|\omega_1,\ldots,\omega_r).
\end{equation}
The usual gamma function
\begin{equation}
\Gamma(x)=\int_0^{\infty}\frac{dt}{t^{1-x}}e^{-t},\ \ \ \ 
{\rm Re}(x)>0,
\end{equation}
is related with the modified gamma function $\Gamma_1$ by
\begin{equation}
\Gamma(x)=\sqrt{2\pi}\omega^{\frac12-x}\Gamma_1(\omega x|\omega).
\end{equation}
Then the zeta function regularization used in \cite{Benini:2012ui,Doroud:2012xw} is obtained
\begin{equation}
\left[\prod_{n=0}^{\infty}(x+n)\right]_{\mbox{\scriptsize reg}}=\frac{\sqrt{2\pi}}{\Gamma(x)}.
\end{equation}

\subsection{Perturbative partition function of four dimensional gauge theory}\label{subapp:pert_4d}

Using a formal expansion
\begin{equation}
\frac{1}{1-e^x}=\sum_{n=0}^{\infty}e^{nx},
\end{equation}
the perturbative partition function (\ref{class_D5}) is written as \cite{Nekrasov:2003rj, Nakajima:2003uh}
\begin{align}
Z_N^{\mbox{\scriptsize cD5}}(\epsilon_1,\epsilon_2,\vec{a})&=\prod_{\begin{subarray}{c}\alpha,\beta=1\\(\alpha\neq \beta)\end{subarray}}^N
\Gamma_2(a_{\alpha\beta}|\epsilon_1,\epsilon_2)\nonumber\\
&=
\prod_{\begin{subarray}{c}\alpha,\beta=1\\(\alpha\neq \beta)\end{subarray}}^N
\exp\Big(\left.\frac{\partial}{\partial s}\right|_{s=0}\frac{1}{\Gamma(s)}\int_0^{\infty}\frac{dt}{t^{1-s}}\frac{e^{-a_{\alpha\beta}t}}{(e^{-\epsilon_1t}-1)(e^{-\epsilon_2t}-1)}\Big).
\end{align}

\section{Stringy $U(1)$ instanton counting, simple Hurwitz theory, and topological strings}\label{app:st_inst_hurwitz}

In this appendix, we discuss relations between the stringy $U(1)$ instanton counting, the simple Hurwitz theory, and the topological A-model on local toric curve. Let us consider the generating function of the stringy $U(1)$ $k$-instanton partition functions (\ref{sinst}) for $N=1$ on the ($r$-scaled) anti-self-dual Omega background $\epsilon_1=-\epsilon_2=\hbar/r$:\footnote{In this background, as shown in (\ref{sinst_ASD}),
the stringy instanton partition function does not have the $\alpha'$ corrections, and thus (\ref{sinst_classic}) for $N=1$ is obtained as the exact result.}
\begin{equation}
Z_{\scriptsize U(1)}^{\mbox{\scriptsize SI}}(\hbar,z,\mathfrak{q})\equiv 1+\sum_{k=1}^{\infty}\big((z\overline{z})^{-ira}\mathfrak{q}^2\big)^kZ_{k,1}\left(\hbar/r,-\hbar/r,a,z\right)=\sum_{\mu}\Big(\frac{\mathfrak{q}}{\hbar}\Big)^{2|\mu|}
\frac{(z\overline{z})^{i\hbar\sum_{s\in \mu}c_{\mu}(s)}}{\prod_{s\in \mu}h_{\mu}(s)^2},
\label{U1_string_inst}
\end{equation}
where $c_{\mu}(s)=\sum_{(i,j)\in\mu}(j-i)$ and $h_{\mu}(s)=a_{\mu}(s)+\ell_{\mu}(s)+1$ is the hook length. By expanding this partition function around $\hbar=0$:
\begin{equation}
Z_{\scriptsize U(1)}^{\mbox{\scriptsize SI}}(\hbar,z,\mathfrak{q})=\sum_{k,\ell=0}^{\infty}(i\hbar\log z\overline{z})^{2\ell-2k}
\frac{(i\mathfrak{q}\log z\overline{z})^{2k}}{(2\ell)!}G_{k,\ell},
\end{equation}
one finds that the coefficients
\begin{equation}
G_{k,\ell}=\sum_{|\mu|=k}\frac{1}{\prod_{s\in \mu}h_{\mu}(s)^2}\Big(\sum_{s\in \mu}c_{\mu}(s)\Big)^{2\ell}
\end{equation}
coincide with the equivariant classical intersection numbers $\langle D^{2\ell}\rangle^{\mbox{\scriptsize cl}}\big|_{\epsilon=0}$ of the divisor class on the Hilbert scheme of points $\mbox{Hilb}_k({\IC}^2)$ on ${\IC}^2$ as computed in (\ref{class_intN1}). One also finds that
\begin{equation}
H_{g,k}^{\scriptsize{\IP}^1}=G_{k,g-1+k}
\label{U1_inst_disHur}
\end{equation}
gives the disconnected simple Hurwitz number of ${\IP}^1$ which counts the degree $k$ ramified cover $f:\ \Sigma_g\ \to\ {\IP}^1$ with $m=2g-2+2k$ simple branch points, where $\Sigma_g$ is a genus $g$ Riemann surface \cite{Hurwitz} (see also e.g. \cite{OkPan}).\footnote{The simple branch point is a branch point such that the branching number is one, and the number $m$ of simple branch points is determined by the Riemann-Hurwitz formula $2g-2+2k=\sum_{i=1}^m(k-\ell_{\mu^i})$, where $\mu^i=(\mu^i_1\ge\mu^i_2\ge\cdots\ge\mu^i_{\ell_{\mu^i}}>0)$ is a profile over an $i$-th branch point. Here the profile of the simple branch point is given by $\mu=(2,1^{k-2})$. Note that the genus $g$ of the disconnected simple Hurwitz numbers can be a negative integer.} Then the connected simple Hurwitz numbers $H_{g,k}^{\scriptsize{\IP}^1\bullet}$ are obtained from the genus expansion of the free energy
\begin{equation}
{\cal F}_{\scriptsize U(1)}^{\mbox{\scriptsize SI}}(\widehat{\hbar},x)=\log Z_{\scriptsize U(1)}^{\mbox{\scriptsize SI}}(\hbar,z,\mathfrak{q})=
\sum_{g=0}^{\infty}\widehat{\hbar}^{2g-2}{\cal F}_g(x)=\sum_{g=0}^{\infty}\widehat{\hbar}^{2g-2}\sum_{k=1}^{\infty}\frac{x^k}{(2g-2+2k)!}H_{g,k}^{\scriptsize{\IP}^1\bullet},
\end{equation}
where $\widehat{\hbar}=i\hbar\log z\overline{z}$ and $x=(i\mathfrak{q}\log z\overline{z})^2$. We see that the perturbative free energies ${\cal F}_g(x)$ are given as \cite{GouJack, GouJackVak}
\begin{align}
&
{\cal F}_0(x)=\sum_{k=1}^{\infty}\frac{k^{k-3}}{k!}x^k,\\
&
{\cal F}_1(x)=-\frac{1}{24}\big(\log(1-y)+y\big),\\
&
{\cal F}_{g\ge 2}(x)=\frac{y^2}{(1-y)^{5(g-1)}}\sum_{i=1}^{3g-5}c_{g,i}y^i,
\end{align}
where $c_{g,i}$ are constants,\footnote{For example these are given by
$$
{\cal F}_2(x)=\frac{y^2}{1440(1-y)^{5}}(6y+1),\ \ {\cal F}_3(x)=\frac{y^2}{725760(1-y)^{10}}(720y^4+3816y^3+3482y^2+548y+9).
$$} and $y=-W(-x)$ is the Lambert $W$ function defined by the inverse function of the spectral curve $x=ye^{-y}$ which has the series expansion $y=\sum_{k=1}^{\infty}\frac{k^{k-1}}{k!}x^k$. The relation between the Hurwitz numbers of ${\IP}^1$ and the intersection numbers on $\mbox{Hilb}_k({\IC}^2)$ was discussed in \cite{LiQinWang}.

\begin{figure}[t]
 \begin{center}
  \includegraphics[width=70mm]{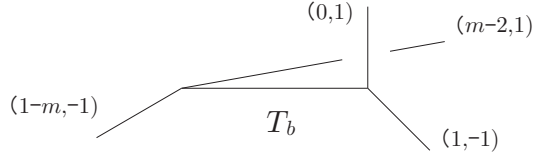}
 \end{center}
 \caption{The toric diagram of the local curve $X_{1,m}={\cal O}(m-2)\oplus{\cal O}(-m)\to {\IP}^1$ (total space of rank two vector bundle over ${\IP}^1$), where $T_b$ is the K\"ahler modulus of ${\IP}^1$.}
\label{local_curve}
\end{figure}

As mentioned in Section \ref{subsec:rel_top_st}, the stringy $U(1)$ instanton partition function (\ref{U1_string_inst}) is obtained from the A-model partition function $Z_{\scriptsize X_{1,m}}^{\mbox{\scriptsize topA}}$ on the local curve $X_{1,m}={\cal O}(m-2)\oplus{\cal O}(-m)\to {\IP}^1$ described in Figure \ref{local_curve}. By the geometric engineering $e^{-T_b}\sim (\beta\mathfrak{q})^2, g_s\sim\hbar$, and the four dimensional limit (\ref{4d_lim}): $\beta \to 0$, $m\to \infty$ with fixed $\beta m=r\log z\overline{z}$, we have
\begin{equation}
Z_{\scriptsize X_{1,m}}^{\mbox{\scriptsize topA}}(g_s,T_b)\ \longrightarrow \ Z_{\scriptsize U(1)}^{\mbox{\scriptsize SI}}(\hbar,z,\mathfrak{q}).
\end{equation}
Therefore one also obtains the relation between the topological A-model on $X_{1,m}$ and the simple Hurwitz theory which was previously discussed in \cite{Caporaso:2006gk}. As pointed out in \cite{Caporaso:2006gk}, the simple Hurwitz theory is related with the $U(1)$ instanton counting with the first and second Casimir operators (see also \cite{Losev:2003py}).\footnote{The (simple) Hurwitz theory is also related with the topological A-model on ${\IP}^1$ \cite{OkPan, LiQinWang}.} Then we see that the stringy $U(1)$ instanton partition function (\ref{U1_string_inst}) also coincides with the $U(1)$ instanton partition function with the first and second Casimir operators.

\section{Equivariant correlators on $\mbox{Hilb}_k({\IC}^2)$}\label{app:equiv_hilb}

Let $\mbox{Hilb}_k({\IC}^2)$ be the Hilbert scheme of points on ${\IC}^2$:
\begin{equation}
\mbox{Hilb}_k({\IC}^2)=\{{\cal J} \subset {\IC}[x,y] | {\cal J}\ \mbox{is an ideal},\ \dim_{{\IC}}{\IC}[x,y]/{\cal J}=k\}.
\end{equation}
The $T$-equivariant cohomology $H_T^*\big(\mbox{Hilb}_k({\IC}^2),{\IQ}\big)$ of $\mbox{Hilb}_k({\IC}^2)$ with the equivariant action $T=U(1)^2$ on ${\IC}^2$ has the Fock space description over ${\IQ}$ as was constructed by Grojnowski and Nakajima \cite{Grojnowski, NakHilb, NakajimaLec} (see also \cite{OkounkovPandharipande}). The Fock module over the Heisenberg algebra $\{\alpha_{\pm n}, n\in{\IN}\ |\ [\alpha_m,\alpha_n]=m\delta_{m+n,0}\}$ is given as follows. The Fock vacuum $|\emptyset \rangle$ is annihilated by $\alpha_{n>0}$: $\alpha_n|\emptyset \rangle=0$ for $n>0$, and the basis of the Fock space ${\cal F}$ is created by $\alpha_{n<0}$:
\begin{equation}
|Y\rangle=\frac{1}{|\mbox{Aut}(Y)|\prod_iY_i}\prod_i \alpha_{-Y_i}|\emptyset \rangle.
\end{equation}
Here $Y$ is a partition with $Y_1\ge Y_2 \ge \cdots \ge Y_{\ell_Y}>0$, and $\mbox{Aut}(Y)$ is the order of the automorphism group of the partition. Then a Fock module $|Y\rangle$ with $|Y|=\sum_{i=1}^{\ell_Y}Y_i=k$ gives an element of $H_T^{2k-2\ell_Y}\big(\mbox{Hilb}_k({\IC}^2),{\IQ}\big)$, and a canonical isomorphism
\begin{equation}
{\cal F}\otimes_{\IQ}{\IQ}[\epsilon_1,\epsilon_2]\cong \bigoplus_{k\ge 0}H_T^*\big(\mbox{Hilb}_k({\IC}^2),{\IQ}\big)
\label{Fock_Coh}
\end{equation}
is obtained, where $\epsilon_{1,2}$ are the equivariant parameters. The inner product on the Fock space which gives the equivariant two-point functions on $\mbox{Hilb}_k({\IC}^2)$ is normalized as
\begin{equation}
\langle Y|Y'\rangle=\frac{(-1)^{|Y|-\ell_Y}}{(\epsilon_1\epsilon_2)^{\ell_Y}|\mbox{Aut}(Y)|\prod_iY_i}\delta_{YY'}.
\label{inner_norm_hilb}
\end{equation}
The Poincar\'e dual of the divisor class in $H_T^*\big(\mbox{Hilb}_k({\IC}^2),{\IQ}\big)$ is given by $|D\rangle=-|2,1^{k-2}\rangle$. The operator of small quantum multiplication by $D$ is given by the $q$-deformed Calogero-Sutherland Hamiltonian \cite{OkounkovPandharipande}
\begin{equation}
H_D=\frac{\epsilon}{2}\sum_{n=2}^{\infty}\Big(\frac{1-q}{1+q}-\frac{1+(-q)^n}{1-(-q)^n}n\Big)\alpha_{-n}\alpha_n+\frac12\sum_{m,n=1}^{\infty}\big(\epsilon_1\epsilon_2\alpha_{-m}\alpha_{-n}\alpha_{m+n}-\alpha_{-m-n}\alpha_m\alpha_n\big),
\label{q_multiple_hilb}
\end{equation}
where $\epsilon=\epsilon_1+\epsilon_2$.

Using this Fock space description, let us compute the equivariant classical intersection numbers of the divisor class in $H_T^*\big(\mbox{Hilb}_k({\IC}^2),{\IQ}\big)$. Let $H_D^{\mbox{\scriptsize cl}}=H_D|_{q=0}$ be the classical part of the operator $H_D$. By $\alpha_n^{\ell}\alpha_{-n}^{k}|\emptyset \rangle=k(k-1)\cdots(k-\ell+1)n^{\ell}\alpha_{-n}^{k-\ell}|\emptyset \rangle$ for $\ell\le k$, one obtains
\begin{align}
H_D^{\mbox{\scriptsize cl}}\alpha_{-2}\alpha_{-1}^{k-2}|\emptyset \rangle
&=
-\epsilon\alpha_{-2}\alpha_{-1}^{k-2}|\emptyset \rangle+\epsilon_1\epsilon_2\alpha_{-1}^k|\emptyset \rangle-\frac12(k-2)(k-3)\alpha_{-2}^2\alpha_{-1}^{k-4}|\emptyset \rangle\nonumber\\
&
-2(k-2)\alpha_{-3}\alpha_{-1}^{k-3}|\emptyset \rangle,
\end{align}
and
\begin{align}
(H_D^{\mbox{\scriptsize cl}})^2\alpha_{-2}\alpha_{-1}^{k-2}|\emptyset \rangle
&=\big(\epsilon^2-\frac12(3k^2+k-12)\epsilon_1\epsilon_2\big)\alpha_{-2}\alpha_{-1}^{k-2}|\emptyset \rangle-\epsilon_1\epsilon_2\epsilon\alpha_{-1}^k|\emptyset \rangle\nonumber\\
&
+\frac32(k-2)(k-3)\epsilon\alpha_{-2}^2\alpha_{-1}^{k-4}|\emptyset \rangle+8(k-2)\epsilon\alpha_{-3}\alpha_{-1}^{k-3}|\emptyset \rangle\nonumber\\
&
+\frac14(k-2)(k-3)(k-4)(k-5)\alpha_{-2}^3\alpha_{-1}^{k-6}|\emptyset \rangle\nonumber\\
&
+3(k-2)(k-3)(k-4)\alpha_{-3}\alpha_{-2}\alpha_{-1}^{k-5}|\emptyset \rangle
+8(k-2)(k-3)\alpha_{-4}\alpha_{-1}^{k-4}|\emptyset \rangle.
\end{align}
Then on $\mbox{Hilb}_k({\IC}^2)$, the equivariant classical intersection numbers of the divisor class are computed as
\begin{align}
\langle D^0\rangle^{\mbox{\scriptsize cl}}&=\langle D^0\rangle=\frac{1}{k!(\epsilon_1\epsilon_2)^k},\ \ \ \ 
\langle D^1\rangle^{\mbox{\scriptsize cl}}=\langle D^1\rangle=0,\ \ \ \ 
\langle D^2\rangle^{\mbox{\scriptsize cl}}=\langle D^2\rangle=-\frac{1}{2(k-2)!(\epsilon_1\epsilon_2)^{k-1}},\nonumber\\
\langle D^3\rangle^{\mbox{\scriptsize cl}}&=\langle D|H_D|D\rangle^{\mbox{\scriptsize cl}}=\frac{\epsilon}{2(k-2)!(\epsilon_1\epsilon_2)^{k-1}},\nonumber\\
\langle D^4\rangle^{\mbox{\scriptsize cl}}&=\langle D|H_D^2|D\rangle^{\mbox{\scriptsize cl}}=
-\frac{\epsilon^2}{2(k-2)!(\epsilon_1\epsilon_2)^{k-1}}+
\frac{3k^2+k-12}{4(k-2)!(\epsilon_1\epsilon_2)^{k-2}},\nonumber\\
\langle D^5\rangle^{\mbox{\scriptsize cl}}&=\langle D|H_D^3|D\rangle^{\mbox{\scriptsize cl}}=
\frac{\epsilon^3}{2(k-2)!(\epsilon_1\epsilon_2)^{k-1}}-\frac{\big(18(k-2)+2(k-2)(k-3)+(3k^2+k-12)\big)\epsilon}{2(k-2)!(\epsilon_1\epsilon_2)^{k-2}},\nonumber\\
\langle D^6\rangle^{\mbox{\scriptsize cl}}&=\langle D|H_D^4|D\rangle^{\mbox{\scriptsize cl}}=
-\frac{\epsilon^4}{2(k-2)!(\epsilon_1\epsilon_2)^{k-1}}\nonumber\\
&\hspace{6.5em}
+\frac{\big(180(k-2)+16(k-2)(k-3)+3(3k^2+k-12)\big)\epsilon^2}{4(k-2)!(\epsilon_1\epsilon_2)^{k-2}}\nonumber\\
&\hspace{6.5em}
-\frac{15k^4+30k^3-105k^2-700k+1344}{8(k-2)!(\epsilon_1\epsilon_2)^{k-3}}.
\label{class_intN1}
\end{align}
By $\epsilon_{1,2}\to -\epsilon_{1,2}$, these results coincide with the classical part of the stringy $U(1)$ instanton counting computed in (\ref{N1stringy_count}). Especially from (\ref{N1_Gk_classic}),
one finds a formula of the classical intersection numbers at $\epsilon=\epsilon_1+\epsilon_2=0$:
\begin{equation}
\langle D^{2\ell}\rangle^{\mbox{\scriptsize cl}}\Big|_{\epsilon=0}=\frac{1}{(\epsilon_1\epsilon_2)^{k-\ell}}\sum_{|\mu|=k}\frac{1}{\prod_{s\in \mu}h_{\mu}(s)^2}\Big(\sum_{s\in \mu}c_{\mu}(s)\Big)^{2\ell},
\end{equation}
where $c_{\mu}(s)=\sum_{(i,j)\in\mu}(j-i)$ and $h_{\mu}(s)=a_{\mu}(s)+\ell_{\mu}(s)+1$ is the hook length defined for a Young diagram $\mu$ as described in Figure \ref{YoungNek} of Section \ref{subsec:n22_glsm}.

The (quantum) equivariant three-point functions in $H_T^*\big(\mbox{Hilb}_k({\IC}^2),{\IQ}\big)$ are also computed as \cite{Bonelli:2013rja}:
\begin{align}
&
\langle D|H_D|D\rangle
=\epsilon\Big(\frac{1-q}{1+q}-\frac{2(1+q^2)}{1-q^2}\Big)\langle D|D\rangle
=-\frac{\epsilon}{(k-2)!(\epsilon_1\epsilon_2)^{k-1}}\Big(\frac12+L_0(q)\Big),
\label{DDD}
\\
&
\langle Y_2^{(1)}|H_D|Y_2^{(1)}\rangle
=-\frac{3}{2}\epsilon\Big(\frac{1-q}{1+q}-\frac{3(1-q^3)}{1+q^2}\Big)\langle Y_2^{(1)}|Y_2^{(1)}\rangle
=\frac{\epsilon}{(k-3)!(\epsilon_1\epsilon_2)^{k-2}}\Big(1+M_0(q)\Big),
\label{DY1Y1}
\\
&
\langle Y_2^{(2)}|H_D|Y_2^{(2)}\rangle
=-2\epsilon\Big(\frac{1-q}{1+q}-\frac{2(1+q^2)}{1-q^2}\Big)\langle Y_2^{(2)}|Y_2^{(2)}\rangle
=\frac{\epsilon}{2(k-4)!(\epsilon_1\epsilon_2)^{k-2}}\Big(\frac12+L_0(q)\Big),
\label{DY2Y2}
\end{align}
where we have changed the equivariant parameters as $\epsilon_{1,2}\to -\epsilon_{1,2}$. Here $|Y_2^{(1)}\rangle=|3,1^{k-3}\rangle$, $|Y_2^{(2)}\rangle=|2^2,1^{k-4}\rangle$ are the cohomological classes dual to (complex) codimension two cycle classes, and
\begin{equation}
L_0(q)=\frac{q}{1-q},\ \ \ \ 
M_0(q)=\frac{q(1-2q)}{1-q+q^2}.
\label{def_l0_m0}
\end{equation}

\section{Equivariant correlators on ${\cal M}_{k,N}$}\label{app:equiv_adhm}

In this appendix, we discuss equivariant correlators on the ADHM moduli space ${\cal M}_{k,N}$ defined in (\ref{ADHM_moduli}). In \cite{Bar}, by extending the construction of Grojnowski and Nakajima \cite{Grojnowski, NakHilb, NakajimaLec} for ${\cal M}_{k,1}\cong\mbox{Hilb}_k({\IC}^2)$ mentioned in Appendix \ref{app:equiv_hilb}, Baranovsky constructed the action of the Heisenberg algebra on the cohomology of ${\cal M}_{k,N}$. The $T$-equivariant cohomology $\bigoplus_{k\ge 0}H_{T}^*\big({\cal M}_{k,N}\big)$, where $T$ is the $N+2$ dimensional torus with the equivariant parameters $a_1,\ldots, a_N, \epsilon_{1,2}$ as in footnote 6 of Section \ref{subsec:st_corr}, forms a tensor product \cite{Bar, Maulik:2012wi}:
\begin{equation}
\bigoplus_{k\ge 0}H_{T}^*\big({\cal M}_{k,N}\big)=H_{T_i}^*\big({\cal M}_{1}\big)^{\otimes N},\ \ \ \
H_{T_i}^*\big({\cal M}_{1}\big)=\bigoplus_{k\ge 0}H_{T_i}^*\big({\cal M}_{k,1},{\IQ}\big).
\end{equation}
Here $T_i$ is the $1+2$ dimensional torus with the equivariant parameters $a_i$ and $\epsilon_{1,2}$, where the torus $e^{ia_i}$ trivially acts on the $i$-th ${\cal M}_{k,1}$.
The Heisenberg algebra is given by 
$\{\alpha_{\pm n}^{(i)}, n\in{\IN}, i=1,\ldots,N\ |\ [\alpha_m^{(i)},\alpha_n^{(j)}]=m\delta_{m+n,0} \delta_{i,j}\}$, and the generators $\alpha_{\pm n}^{(i)}$ only act on the $i$-th tensor factor.

Let $|\mathsf{1}_k\rangle$ be the element in the tensor product of Fock spaces corresponding to
the cohomological degree 0 element $\mathsf{1}_k$ in
$H_{T}^0\big({\cal M}_{k,N}\big)$, and using the isomorphism (\ref{Fock_Coh}), we consider
\begin{equation}
|Y\rangle=\frac{1}{|\mbox{Aut}(Y)|N\prod_IY_I}\prod_I \beta_{-Y_I}|\mathsf{1}_{k-l}\rangle.
\label{Y_high_rank_k}
\end{equation}
Here $Y$ is a partition with $Y_1\ge Y_2 \ge \cdots \ge Y_{\ell_Y}>0$ and $|Y|=\sum_{I=1}^{\ell_Y}Y_I=l$, $\mbox{Aut}(Y)$ is the order of the automorphism group of the partition, and
\begin{equation}
\beta_{\pm n}=\sum_{i=1}^N\alpha_{\pm n}^{(i)}
\label{Baran_op}
\end{equation}
are the Baranovsky operators \cite{Bar}. The Baranovsky operators act on the cohomology ring of ${\cal M}_{k,N}$ as
\begin{equation}
\beta_{\pm n}:\ \ 
\bigoplus_{k}H_{T}^*\big({\cal M}_{k,N}\big)\ \longrightarrow\ 
\bigoplus_{k}H_{T}^*\big({\cal M}_{k\mp n,N}\big),
\end{equation}
and the action $\beta_{-n}, n>0$ increases the cohomological degree by
\begin{equation}
2nN-2.
\label{Mkn_ber_deg}
\end{equation}
Therefore the module (\ref{Y_high_rank_k}) describes an element of $H_{T}^{2lN-2\ell_Y}\big({\cal M}_{k,N}\big)$. On the other hand, the action
$\beta_{n}, n>0$ decreases the cohomological degree by
$2nN-2$, and thus if $n \ge 2$, or $n=1$ for $N\ge 2$, then $\beta_{n} \mathsf{1}_k=0$ \cite{Maulik:2012wi}. Therefore for $N\ge 2$, we consider the module $|\mathsf{1}_k\rangle$ as a ``vacuum''.
The inner product (two-point function on ${\cal M}_{k,N}$) of $|\mathsf{1}_k\rangle$ is given (defined) by the Nekrasov partition function (\ref{Nekrasov_integ}):
\begin{equation}
\langle\mathsf{1}_k, \mathsf{1}_k\rangle=
\langle\mathsf{1}_k|\mathsf{1}_k\rangle=\int_{{\cal M}_{k,N}}1=Z_{k,N}^{\mbox{\scriptsize Nek}}(\epsilon_1,\epsilon_2,\vec{a}),
\end{equation}
where $1 \in H_{T}^*\big({\cal M}_{k,N}\big)$. As a higher rank generalization of (\ref{inner_norm_hilb}), it is natural to give the normalization of the inner product of $|Y\rangle$ in (\ref{Y_high_rank_k}) as
\begin{equation}
\langle Y|Y'\rangle=\frac{(-1)^{|Y|-\ell_Y}c_Y}{(\epsilon_1\epsilon_2)^{\ell_Y}|\mbox{Aut}(Y)|N \prod_IY_I}\delta_{YY'}\times Z_{k-l,N}^{\mbox{\scriptsize Nek}}(\epsilon_1,\epsilon_2,\vec{a}),
\label{inner_norm_adhm}
\end{equation}
where $c_Y$ is a constant which is not fixed in this paper.
In \cite{Maulik:2012wi}, by Maulik and Okounkov it was shown that the operator of quantum multiplication by the divisor for ${\cal M}_{k,N}$ is given by the $q$-deformed Hamiltonian of a coupled $N$-tuple of Calogero-Sutherland system:\footnote{We have slightly changed the convention of \cite{Maulik:2012wi}.}
\begin{align}
H_D&
=
\frac12\sum_{i=1}^N\sum_{m,n=1}^{\infty}\big(\epsilon_1\epsilon_2\alpha_{-m}^{(i)}\alpha_{-n}^{(i)}\alpha_{m+n}^{(i)}-\alpha_{-m-n}^{(i)}\alpha_m^{(i)}\alpha_n^{(i)}\big)
\nonumber\\
&\ \
+\sum_{i=1}^N\sum_{n=1}^{\infty}\Big(a_i+\frac{\epsilon}{2}(1-n)\Big)\alpha_{-n}^{(i)}\alpha_{n}^{(i)}
-\epsilon \sum_{i<j}\sum_{n=1}^{\infty}n\alpha_{-n}^{(j)}\alpha_{n}^{(i)}
\nonumber\\
&\ \
-\epsilon \sum_{n=1}^{\infty} \frac{q^n}{1-q^n}n \beta_{-n}\beta_n,
\label{q_multiple_adhm}
\end{align}
where $\epsilon=\epsilon_1+\epsilon_2$. In \cite{Maulik:2012wi}, it was also shown that the divisor generates the quantum cohomology ring of ${\cal M}_{k,N}$. Note that in the $N=1$ case, we need to modify the above operator $H_D$ by adding
\begin{equation}
\frac{\epsilon q}{1-q}\sum_{n=1}^{\infty}\alpha_{-n}\alpha_{n},
\end{equation}
and then by putting $a_1=0$, and $q \to -q$, the operator (\ref{q_multiple_hilb}) for $\mbox{Hilb}_k({\IC}^2)$ is obtained.

In what follows, using the last term in (\ref{q_multiple_adhm}):
\begin{equation}
H_D^{\mbox{\scriptsize quant}}\equiv -\epsilon \sum_{n=1}^{\infty} \frac{q^n}{1-q^n}n \beta_{-n}\beta_n,
\label{q_multiple_q_adhm}
\end{equation}
and $\beta_{n}|\mathsf{1}_k\rangle=0$,
we compute the quantum parts of two three-point functions related with the computations in Section \ref{sec:quant_st_inst}:
\begin{equation}
\langle D, Y_p, Y_p\rangle^{\mbox{\scriptsize qu}}\equiv
\langle Y_p|H_D^{\mbox{\scriptsize quant}}|Y_p\rangle.
\end{equation}
Here $Y_1\equiv \beta_{-1}\mathsf{1}_{k-1}\in H_{T}^{2N-2}\big({\cal M}_{k,N}\big)$, $Y_2\equiv \beta_{-1}^2\mathsf{1}_{k-2}\in H_{T}^{4N-4}\big({\cal M}_{k,N}\big)$, and $D$ is the Poincar\'e dual of the divisor class.\footnote{For $N=2$, we see that $Y_1=D$.}
By (\ref{inner_norm_adhm}) and (\ref{q_multiple_q_adhm}), we obtain
\begin{align}
&
\langle D, Y_1, Y_1\rangle^{\mbox{\scriptsize qu}}=-\frac{c_1\epsilon}{\epsilon_1\epsilon_2}L_0(q)Z_{k-1,N}^{\mbox{\scriptsize Nek}}(\epsilon_1,\epsilon_2,\vec{a}),
\label{DY1Y1_adhm}
\\
&
\langle D, Y_2, Y_2\rangle^{\mbox{\scriptsize qu}}=-\frac{c_2\epsilon}{\epsilon_1^2\epsilon_2^2}L_0(q)Z_{k-2,N}^{\mbox{\scriptsize Nek}}(\epsilon_1,\epsilon_2,\vec{a}),
\label{DY2Y2_adhm}
\end{align}
where $c_{1,2}$ are constants, and $L_0(q)$ is defined in (\ref{def_l0_m0}). It would be interesting to determine the classical parts and fix the normalization constants.

\section{Exact K\"ahler potential for Calabi-Yau threefolds and fourfolds}\label{app:exact_kahler}

In this appendix, we summarize the exact K\"ahler potentials on quantum K\"ahler moduli spaces of Calabi-Yau threefolds (e.g. \cite{Jockers:2012dk}) and fourfolds (conjectured in \cite{Honma:2013hma}). The K\"ahler moduli space ${\cal M}_{\mbox{\scriptsize K\"ahler}}(X)$ of Calabi-Yau $d$-fold $X$ is defined by $H^1(\wedge^1T^*X)$, where $T^*X$ is the holomorphic cotangent bundle on $X$. By considering the NLSM propagating on $X$, the K\"ahler moduli space ${\cal M}_{\mbox{\scriptsize K\"ahler}}(X)$ is quantized by the $\alpha'$ corrections. 

For Calabi-Yau threefold, it is known that around a large radius point the quantum-corrected K\"ahler potential $K$ on ${\cal M}_{\mbox{\scriptsize K\"ahler}}(X)$ is given by
\begin{align}
e^{-K}&=
-\frac{i}{3!}\kappa_{ijk}(t^i-{\overline t}^i)(t^j-{\overline t}^j)(t^k-{\overline t}^k)+\frac{1}{4\pi^3}\zeta(3)\chi(X)\nonumber\\
&
-i\Big[\frac{\partial}{\partial t^k}F(t)+\frac{\partial}{\partial \overline{t}^k}\overline{F(t)}\Big](t^k-\overline{t}^k)
+2i\big(F(t)-\overline{F(t)}\big),
\label{CY3_Kahler}
\end{align}
where $i,j,k=1,\ldots,h^{1,1}(X)$. Here $t^{\ell}$ are the complexified K\"ahler parameters, $\kappa_{\ell mn}$ are the classical triple intersection numbers of divisors on $X$, $\chi(X)$ is the Euler characteristic of $X$, and
\begin{equation}
F(t)=\frac{1}{(2\pi i)^3}\sum_{\beta\in H_2(X,{\IZ})\setminus \{0\}}n_{\beta}{\rm Li}_3(q^{\beta}),\ \ \ \ 
q^{\beta} \equiv e^{2\pi i\beta_kt^k}
\end{equation}
is the prepotential which gives the Gromov-Witten invariants $n_{\beta}$ defined by the holomorphic maps $\partial\phi=0$ in the (A-twisted) NLSM $\phi: {\IP}^1\to X$. Let ${\cal O}_{J_i}$ be the observables associated with $J_{i}\in H^{1,1}(X)$, then the prepotential $F(t)$ can be obtained from the three-point function $\langle{\cal O}_{J_i}{\cal O}_{J_j}{\cal O}_{J_k}\rangle$ on ${\IP}^1$ in the topological A-model via the relation
\begin{equation}
\langle {\cal O}_{J_i}{\cal O}_{J_j}{\cal O}_{J_k}\rangle=\kappa_{ijk}+\frac{\partial^3}{\partial t^i\partial t^j\partial t^k}F(t).
\end{equation}

For Calabi-Yau fourfold, it was conjectured that the K\"ahler potential $K$ around a large radius point is given by \cite{Honma:2013hma}
\begin{align}
e^{-K}&=
\frac{1}{4!}\kappa_{ijk\ell}(t^i-\overline{t}^i)(t^j-\overline{t}^j)(t^k-\overline{t}^k)(t^{\ell}-\overline{t}^{\ell})+\frac{i}{4\pi^3}\zeta(3)C_{\ell}(t^{\ell}-\overline{t}^{\ell})\nonumber\\
&
+\frac12\big(G_{k\ell}(t)+\overline{G_{k\ell}(t)}\big)(t^k-\overline{t}^k)(t^{\ell}-\overline{t}^{\ell})\nonumber\\
&
-\big(H_{\ell}(t)-\overline{H_{\ell}(t)}\big)(t^{\ell}-\overline{t}^{\ell})
+\frac12\eta^{mn}\big(F_{mn;\ell}(t)-\overline{F_{mn;\ell}(t)}\big)(t^{\ell}-\overline{t}^{\ell})\nonumber\\
&
-\frac12\eta^{mn}\big(F_m(t)-\overline{F_m(t)}\big)\big(F_n(t)-\overline{F_n(t)}\big),
\label{CY4_Kahler}
\end{align}
where $i,j,k,\ell=1,\ldots,h^{1,1}(X)$ and $m,n=1,\ldots,h^{2,2}_{\mbox{\scriptsize prim}}(X)$.\footnote{$h^{2,2}_{\mbox{\scriptsize prim}}(X)$ is the dimension of the primary subspace $H^{2,2}_{\mbox{\scriptsize prim}}(X)\subset H^{2,2}(X)$ whose elements are obtained from the wedge products of the elements of $H^{1,1}(X)$.} Here $\kappa_{ijk\ell}$ are the classical quadruple intersection numbers of divisors on $X$, $C_{\ell}=\int_Xc_3(X)\wedge J_{\ell}$ defined by the third Chern class $c_3(X)$ of $X$ and $J_{\ell}\in H^{1,1}(X)$, $\eta^{mn}$ is the inverse matrix of the intersection matrix $\eta_{mn}=\int_XH_m\wedge H_n$ on $H^{2,2}_{\mbox{\scriptsize prim}}(X)$. Similar to the case of Calabi-Yau threefold, the quantum corrections are given by the generating functions
\begin{equation}
F_n(t)=\frac{1}{(2\pi i)^2}\sum_{\beta\in H_2(X,{\IZ})\setminus \{0\}}n_{\beta,n}{\rm Li}_2(q^{\beta})
\end{equation}
which give the Gromov-Witten invariants $n_{\beta,n}$ defined by the holomorphic maps $\partial\phi=0$ intersecting with the cycle dual to $H_n\in H^{2,2}_{\mbox{\scriptsize prim}}(X)$ in the (A-twisted) NLSM $\phi: {\IP}^1\to X$. The generating function $F_n(t)$ is obtained from the three-point function $\langle{\cal O}_{J_i}{\cal O}_{J_j}{\cal O}_{H_n}\rangle$ on ${\IP}^1$ in the topological A-model via the relation
\begin{equation}
\langle {\cal O}_{J_i}{\cal O}_{J_j}{\cal O}_{H_n}\rangle=\kappa_{ijn}+\frac{\partial^2}{\partial t^i\partial t^j}F_n(t),
\end{equation}
where ${\cal O}_{H_n}$ is the observable associated with $H_n\in H^{2,2}_{\mbox{\scriptsize prim}}(X)$, and $\kappa_{ijn}=\int_XJ_i\wedge J_j\wedge H_n$ is the classical intersection number. In the conjectural formula (\ref{CY4_Kahler}), the generating functions
\begin{equation}
G_{k\ell}(t)=\frac{1}{(2\pi i)^2}\sum_{\beta\in H_2(X,{\IZ})\setminus \{0\}}n_{\beta,k\ell}{\rm Li}_2(q^{\beta})
\end{equation}
count the Gromov-Witten invariants $n_{\beta,k\ell}$ defined by the holomorphic maps $\partial\phi=0$ intersecting with the cycle dual to $J_k\wedge J_{\ell}\in H^{2,2}_{\mbox{\scriptsize prim}}(X)$, and by definition these generating functions are written by a linear combination of $F_n(t)$. Other quantities in (\ref{CY4_Kahler}) are defined by
\begin{align}
&
H_{\ell}(t)=\int^{t^{\ell}}_{i\infty}G_{\ell\ell}(t)dt^{\ell}+2\sum_{k\neq \ell}\int^{t^k}_{i\infty}G_{k\ell}(t)dt^k\bigg|_{\scriptsize t^{\ell}=i\infty},\\
&
F_{mn;\ell}(t)=\int^{t^{\ell}}_{i\infty}\partial_{\ell}F_m(t)\partial_{\ell}F_n(t)dt^{\ell}.
\end{align}


\end{document}